\documentclass[usenatbib,letterpaper]{mn2e}
\usepackage{graphicx,epsfig,amsmath,soul}
\usepackage[normalem]{ulem}

\newcommand{\ltaraw}{$\; \buildrel < \over \sim \;$}
\newcommand{\lta}{\lower.5ex\hbox{\ltaraw}}
\newcommand{\gtaraw}{$\; \buildrel > \over \sim \;$}
\newcommand{\gta}{\lower.5ex\hbox{\gtaraw}}

\title[Investigating the Andromeda Stream: I]
  {Investigating the Andromeda Stream:  I. Simple
  Analytic Bulge-Disk-Halo Model for M31}
\author[J. Geehan et al.]
  {J. J.~Geehan$^1$\thanks{E-mail: jgeehan@uvic.ca (JJG), fardal@fcrao1.astro.umass.edu (MF), babul@uvic.ca (AB), raja@ucolick.org (PG)}, M. A. Fardal$^{1,2}$, A. Babul$^1$,
  P.~Guhathakurta$^3$ \\
  $^1$Dept. of Physics \& Astronomy, University of Victoria,
    Elliott Building, 3800 Finnerty Rd., Victoria, BC V8P 1A1, Canada\\
  $^2$Dept.\ of Astronomy, University of Massachusetts,
        Amherst, MA 01003, USA\\
  $^3$UCO/Lick Observatory, Dept.\ of Astronomy \& Astrophysics,
        Univ. of California, 1156 High St., Santa Cruz, CA 95064, USA}

\date{Accepted 2005 November 10. Received 2005 October 30; in original form 2005 January 13}

\pagerange{\pageref{firstpage}--\pageref{lastpage}} \pubyear{2006}

\def\msun{{\rm\; M}_\odot}
\def\kpc{{\rm\; kpc}}

\begin{document}

\label{firstpage}

\maketitle

\begin{abstract}
This paper is the first in a series which studies interactions between
M31 and its satellites, including the origin of the giant southern stream.
We construct accurate yet simple analytic models for the potential of the M31 
galaxy to provide an easy basis for calculation of orbits in M31's halo.  We 
use an NFW dark halo, an exponential disk, a Hernquist bulge, and a central 
black hole point mass to describe the galaxy potential.  We constrain the 
parameters of these functions by comparing to existing surface brightness, 
velocity dispersion, and rotation curve measurements of M31.  Our description 
provides a good fit to the observations, and  agrees well with more
sophisticated modeling of M31.  While in many respects the parameter
set is well constrained, there is substantial uncertainty in the outer
halo potential and a near-degeneracy between the disk and halo
components, producing a large, nearly two-dimensional allowed region in
parameter space.  We limit the allowed region using theoretical expectations 
for the halo concentration, baryonic content, and stellar $M/L$ ratio, finding a
smaller region where the parameters are physically plausible.  
Our proposed mass model for M31 has 
$M_{\rm bulge}=3.2\times 10^{10}\msun$, $M_{\rm disk}=7.2\times 10^{10}\msun$, 
and $M_{200}=7.1\times 10^{11}\msun$, with uncorrected (for internal and foreground
extinction) mass-to-light ratios of $M/L_R=3.9$ and $3.3$ for the bulge and disk,
respectively.  We present some
illustrative test particle orbits for the progenitor of the stellar stream 
in our galaxy potential, highlighting the effects of the remaining uncertainty
in the disk and halo masses.
\end{abstract}

\begin{keywords}
galaxies: M31 -- galaxies: kinematics and dynamics
\end{keywords}

\section{Introduction}
The detritus resulting from the disruption and assimilation of a satellite
galaxy by its larger host is a veritable treasure trove of clues and information
about the galaxy assembly process, the nature of the merging subunits, as well as 
the dynamical properties of the parent system.  Typically in the form of stellar 
streams, this detritus can remain spatially and kinematically coherent in the 
halo of the larger galaxy for several billion years \citep{joh96,hel99}.
Identifying and quantifying these features in detail, however, is a challenging task as 
it often entails resolving and measuring properties of individual stars in the 
stellar halo of the galaxies.   At present, such studies are only feasible 
for the galaxies in the Local Group.   The Milky Way stellar halo has been 
the focus of detailed scrutiny for many years now resulting in the detection 
of a number of coherent features in the star counts 
(see \citealp{yan03, maj04, mart04, law05} and references therein).
Unfortunately, as noted by \citet{new02}, our vantage point within the 
Galaxy greatly complicates the interpretation of these features.
Consequently, the recent uncovering of rich substructure \citep[][]{iba01,
fer02,mcc03,morr03,merr03,fer04,guh04,zuc04,merr04}
in the stellar halo of the Andromeda galaxy (M31) is highly tantalizing.

The most striking of these features is the giant southern stellar
stream extending out from the south-eastern part of M31's disk. First
reported by \citet{iba01}, the stream stars have since been targeted
for careful photometric and spectroscopic analyses resulting in the
determination of the distance to the stream at various locations along
its length \citep{mcc03} as well as the measurement of stellar
kinematics in a number of locations \citep[][]{iba04, guh04}.
Jointly, these observations indicate that the giant southern stream
extends away from us below the disk of M31, out to a distance
of $\sim 100$ kpc away from the M31 center.  There is also
a strong velocity gradient along the
body of the stream, with the outer regions being at rest with respect
to M31 while the inner regions approach us at $300$ km s$^{-1}$ with
respect to M31.

Intense observational effort notwithstanding, the source of the 
giant southern stream has yet to be identified.  We do not know whether
the progenitor survives and if so, where it is, although several faint 
features in the inner halo of M31, such as the newly detected satellite
And VIII \citep{morr03}, have been touted
as possibilities.  One approach to identifying the progenitor of the
stream, or at least its fate, is to reconstruct its orbit using the 
stream properties as constraints.  Both \citet{iba04} and \citet{fon04}
have initiated  efforts in this direction.  As we shall 
illustrate, both in this and in Paper II \citep{paperII}, 
a detailed study of the stream dynamics and of its progenitor orbit, 
however, requires a realistic, reasonably accurate model for the potential 
of M31 that is also preferably easy to use and straightforward
to alter for experimentation purposes.

The simplest approach to describing the potential of M31 is by modeling
the mass distribution in the galaxy.  The earliest efforts at doing so date 
back to \citet{bab38}, 
followed by a series of investigations by 
\citet{wy42}, \citet{kuz43},
\citet{kuz52}, \citet{schw54}, and \citet{sch57} over the course
of the next two decades.
Subsequent improvement in the quality of photometric and spectroscopic 
data in the 70s and the 80s led to a reassessment of the mass models
by \citet{deh75}, \citet{monn77}, \citet{sim79},
and  \citet{ken89}.  As heroic as these efforts were, these models were
constructed at a time when the shape and the parameters describing 
dark halos were largely unknown and unconstrained.  Most recently,
the problem has been revisited by \citet{klyp02}, \citet{wid03} and
\citet{wid05}.  However, the mass models put forth by these
groups, though sophisticated, are neither simple nor easy to use
especially for the purposes of orbit calculations.  The \citet{wid03} 
mass models, for example, are specified in terms of a set of distribution 
functions. While these distribution functions can be used to compute the mass
density distribution for M31 as well as the associated gravitational 
potential, the derivation is implicit and must be solved for iteratively.
In other words, a closed-form analytic description of the density and the
potential is not available.  Similarly, the Klypin et al.~model is also
not available in closed analytic form.

In this paper, we attempt to remedy this by presenting a reasonably 
accurate yet simple analytic description of the M31 mass distribution.
Our principal aim is to arrive at
a description of the gravitational potential that is both suitable for the 
purposes of computing satellite orbits and sufficiently transparent that
it can be easily altered for the sake of experimentation.
We achieve this by decomposing the mass distribution in M31 into four 
components: the central black hole (BH), the bulge, the disk and the 
extended halo.   The components are
modelled using well-known functional forms whose associated
potentials are easy to compute.   The appropriate values of the associated 
structural parameters are determined by requiring that the model
is in agreement with the observed M31 rotation, surface brightness, 
and bulge velocity dispersion profiles.  In \S~\ref{m31pot}, we 
describe the analytic functional forms we have chosen for the four components.
In \S~\ref{m31obs}, we review the M31 observations used to constrain
the structural parameters, and visually compare
to our best-fit model.  
In \S~\ref{modelbehave}, we examine
the allowed region of parameter space, which shows both strong 
constraints in some directions and near-degeneracies in others.
We discuss additional physical constraints on the solutions
besides those used in the fit.  
Two ``constrained best-fit'' solutions resulting from this discussion
are presented along with those of
our formal best-fit solution in Table~\ref{param};
we direct readers interested only in the final product to this table.
In \S~\ref{obscon},
we briefly illustrate the relevance of our present efforts, 
by computing sample orbits for the progenitor
of the giant southern stream.  
We discuss 
the effect of the uncertainty in the potentials on the orbits,
and compare our orbits to those
obtained in some simple analytic potentials that have been recently
used for this purpose.
A summary of our key results is presented in 
\S~\ref{summary}.   
For completeness, we note that our adopted
cosmology is a spatially flat $\Lambda$CDM universe with 
$\Omega_m =0.14h^{-2}$, $\Omega_{b}=0.024h^{-2}$, and a Hubble constant 
of $h=0.71$ in units of 100 km/s/Mpc. 

\section{Components of our M31 Mass Model}\label{m31pot}

In principle, a detailed mass model of M31 ought to comprise of several 
components:  the central black hole, the nucleus, the bulge, the bar, 
the spheroid, the thick and thin stellar disks and the thin gaseous disk.  
In the interest of simplicity, we restrict ourselves to only four components: 
the central BH, the bulge, the disk and the extended halo.  Moreover,
our use of simple analytic functions to describe these components
necessitates making some simplifying assumptions.  We discuss these
as well as our choice of functional forms below.

\subsection{The Central Black Hole}\label{bhpot}

The very central region of M31 is 
comprised of a central BH
and a distinct small-scale stellar nuclear component that is photometrically and dynamically 
separate from the bulge and the large-scale galactic disk (see \citealp{kor95} and 
references therein).  
The most recent estimate of the BH mass is
$M_{\bullet}=(5.6\pm 0.7)\times 10^7 \;\msun$ \citep{sal04},
while the mass of the stellar component is estimated to be $\lta$2$\times 10^7\;\msun$, 
within $\sim 10$ pc \citep{bac01, pen02, sal04}.

In terms of the total M31 potential, the two nuclear components are  
only relevant on scales $r < 20$ pc and for all practical purposes, can be 
ignored in the context of the calculations of interest to us.   However, to facilate 
wider use of our M31 model, we include the dynamically  more important of the two, 
the BH component, in the form of a point mass located at the center of our galaxy, 
and fix its mass to the value quoted above.
The inclusion of this component is necessary, for example, in order to correctly model 
the velocity components in the nucleus of M31 (cf.  Figure~\ref{veldisplot}).

\subsection{The Galactic Bulge}\label{bulgepot}

We model the bulge of M31 as a spherically symmetric mass distribution represented
by a Hernquist profile \citep{her90}:
\begin{equation} 
\rho_{b}(r)= \left(\frac{M_{b}}{2\pi r_b^3}\right)\frac{1}{(r/r_b)(1+r/r_b)^3},
\label{hern}
\end{equation}
where $M_{b}$ is the total mass of the bulge and $r_{b}$ is its scale radius.
The mass profile and the potential 
corresponding to this density distribution are 
\begin{align}
M_{b}(r) & =\frac{M_{b} r^{2}}{(r_{b}+r)^2}, \\
\Phi_{b}(r) &=-\frac{G M_{b}}{r_{b}+r}. 
\end{align}
We considered using the more general density profile of \citet{deh93} 
where $\rho_b(r)\propto (r/r_b)^{-\gamma} (1+r/r_b)^{\gamma -3}$ but found that 
the results for $0.1 \leq \gamma\leq 2$ were all equally good.  For simplicity, we
have opted to use the $\gamma =1$ Hernquist profile.   We also note that the both the
bulge mass and the bulge mass-to-light ratio are quite insensitive to variations
in $\gamma$ within the range noted.  This gives us confidence that 
the bulge parameters that we quote in Table~\ref{param},
and more generally our bulge-disk decomposition, are robust regardless of
the precise nature of the inner density structure in the bulge.

On the other hand, using a spherical model for the bulge is clearly  a simplification.  
Detailed modeling based on surface photometry suggests that at the very least, 
the bulge ought to be modelled as an oblate spheroid with axis ratio of $\sim 0.8$ 
\citep{ken83, ken89, pen02, wid03} and that in actuality, it is almost
certainly triaxial \citep{lin56, star77, ken89, star94, ber01,  ber02}.
In opting to avoid the complications of using a non-spherical mass
distribution, we are primarily guided by the purpose of this paper,
which is to construct orbits to model the giant southern stream.  This
stream has a pericenter estimated to be at $\sim 2$ to $4$ kpc, and an
apocenter of $\sim 100$ kpc \citep{iba04,fon04}, so it spends
little to no time in a region where the asphericity of the bulge could
have any dynamical effect.  After all, the equipotential surfaces 
tend to spheres at large radii
even if the density distribution does not.  However, 
if our models are used to treat dynamics at smaller radii,
the reader should keep this simplification in mind.

\subsection{The Galactic Disk}\label{diskpot}

In order to model the M31 galactic disk, we begin by assuming, as suggested by the 
observations of \citet{wal87}, that the disk mass distribution can be described by 
an exponential surface density profile:  
\begin{equation}
\Sigma_{d}(R)=\Sigma_{0} e^{-R/R_{d}},
\end{equation}
where $\Sigma_{0}$ is the central surface density, $R_{d}$ is the scale length of the
disk, and $R$ is the distance from the centre of M31 in the plane of the disk.
This corresponds to a disk mass within a sphere of radius $r$ of
\begin{equation}
M_{d}(r)=2\pi \Sigma_{0} R_{d}^2\left[1 - (1+r/R_{d}) e^{-r/R_{d}} \right].
\end{equation}
Here, $r=\sqrt{R^{2}+z^{2}}$ is the distance from the centre of M31 and 
$z$ is the distance perpendicular to the plane of the disk.  

For an infinitesimally thin disk,
\citet{bin87} give the expression for the potential as:
\begin{equation}
\Phi_{d}(R,z)=-2\pi G\Sigma_{0}R_{d}^{2}
\int_{0}^{\infty} \frac{J_{0}(kR)e^{{-k|z|}}dk}{\left[1+(kR_{d})^{2}\right]^{3/2}},
\end{equation}
and in the plane of the disk, this thin disk potential implies a circular 
velocity profile given by \citep[][]{bin87}:
\begin{equation}
\label{eqn.vcflat}
V_{c,d}^{2}(R)=4\pi G \Sigma_{0} R_{d} y^{2}\left[I_{0}(y)K_{0}(y)-I_{1}(y)K_{1}(y)\right],
\end{equation}
where $y=R/2R_{d}$ and ($J_0$, $I_0$, $K_0$, $I_1$, $K_1$) are Bessel functions.

Below, we shall also have occasion to use an extremely simple toy
model that we will refer to as the ``spherical disk'' model.  In this
model, we assume that the disk mass within a sphere of radius $r$ is
distributed in a spherically symmetric fashion rather than
concentrated in a thin disk.  
Then the corresponding gravitational potential is simply
\begin{align}
\label{phisphrdisk}
\Phi_{d,sp}(r) &=-2 \pi G\Sigma_{0}R_{d}^{2} \left[\frac{1-e^{{-r/R_{d}}}}{r}\right] .
\end{align}
Unless otherwise stated, the results
presented in this paper will be based on the thin disk model.

\subsection{The Extended Dark Halo}\label{halopot}

Finally, we consider the extended dark matter halo of M31.  We assume that this
component can be adequately modeled as a spherically symmetric system.  We recognize
that the issue concerning the sphericity of the halo is a contentious one, but 
given that it remains observationally unresolved even in the case of the Milky Way,
we feel it more in keeping with the spirit of our approach to adopt the simplest
model.   To describe the run of density with radius, we adopt the
NFW profile \citep{nav96}:
\begin{equation}
\rho_{h}(r)=\frac{\delta_{c} \rho_{c}}{(r/r_{h})(1+r/r_{h})^2},
\label{nfw}
\end{equation}
where $\rho_{c}=277.72h^2\;\msun/\kpc^2$ is the present-day critical density, $h=0.71$ is the 
Hubble constant in units of 100 km/s/Mpc, $\delta_{c}$ is a dimensionless density 
parameter, and $r_{h}$ is the halo scale radius.   N-body simulations based on 
the hierarchical clustering scenario for structure formation within the cold 
dark matter cosmogony suggest that spherically averaged halo density profiles are 
well described by the above profile. There is much debate over the exact 
exponent of the density in the inner cusp,
but in our specific instance, this will not matter since 
the potential within the region in question will be dominated by the disk 
and/or the bulge components.  

The mass profile and the corresponding potential for an NFW halo are given by:
\begin{align}
M_{h}(r)= & 4\pi \delta_{c} \rho_{c} r_{h}^{3} 
 \left[ \ln\left(\frac{{r+r_h}}{r_h}\right) - \frac{r}{{r+r_h}} \right], \label{nfwmass} \\
\Phi_{h}(r)= & -4\pi G \delta_{c}\rho_{c} r_{h}^{2}\left(\frac{r_h}{r}\right) 
\ln\left[\frac{{r+r_h}}{r_h}\right].
\end{align}

\section{Specifying the M31 Model Parameters}\label{m31obs}

The BH-Bulge-Disk-Halo mass model described above has a total of  6 structural parameters
remaining to be specified: $\delta_c$, $r_h$, $M_b$, $r_b$, 
$\Sigma_0$, and $R_d$, or in other words a scale radius and normalization
for each of the bulge, disk, and halo components.
We constrain these parameters using a number
of extant observations of M31.

Before discussing the data and our efforts to construct the mass model, we
briefly review what is known about the configuration of M31.  The Andromeda
galaxy lies at a distance of 784$\pm$24 kpc \citep{stan98} from the Milky Way,
and has a mean radial velocity of $-300 \pm 4$ km/s \citep{deva91}.
We look at the galaxy from below, as is evident from images of the disk 
dust lanes projected onto the bulge. We take the inclination to be 77$^\circ$. 
This is the generally accepted value, although estimates 
range from 74$^\circ$ to 79$^\circ$ \citep{rub69, wal88, ma97}.
The galaxy spins counterclockwise on the sky.  We take the position angle 
of the disk to be 37$^\circ$.  The position angle of the bulge major axis appears to be offset from 
that of the disk by $\sim$ +10$^\circ$  
to +20$^\circ$ \citep{mce83, kor88}
and additionally, the bulge isophotes have
distinctly ``box-shaped'' appearances with ellipticities that increase with radius.  
These latter features, 
as well as the anomalously high non-circular gas velocities in the inner disk,
have generally been interpreted as signatures of a 
triaxial bulge \citep{lin56,star77, ken89, star94, ber01, ber02}.
In addition, analyses of both the HI and the light distribution in M31 disk
suggest that the disk is significantly warped, especially in the outer regions
(cf.~\citealp{saw82, inn82, wal87, wal88, mor94}).  \citet{brau91} argues that 
in addition to this warping, the inclination of the M31 HI disk also
varies as a function of both radius and azimuth and that it may even be flaring toward
the outer regions.  The latter is not altogether surprising in light of the number
of satellites that appear to be interacting with the galaxy;  the flaring of the disk
is one of the generic features of such interactions (cf. Fardal, Babul \& Barnes, 
in preparation).

Over the past two decades, detailed studies of M31 have generated a large volume of kinematic
and photometric data.   For practical considerations, we restrict ourselves to a
limited number of data sets.  Specifically, we constrain our mass model using
both the M31 major-axis and average surface brightness profiles, the bulge major and
minor axis velocity dispersion profiles, the disk rotation profile, and the dynamically 
derived total mass estimates in the intermediate and outer halo.  These are described 
more fully in \S~\ref{surfbprof}--\ref{massprof}.  
In order to relate our mass model to the light profiles, we are required to 
introduce two additional parameters: $\left(M/L_R\right)_b$ and 
$\left(M/L_R\right)_d$, or the mass-to-light ratios of the bulge and
the disk.  We assume that these two ratios are constant over the entire
bulge and disk; this is probably not exactly correct but should be reasonably
valid, given the modest color gradient in M31 \citep{wal87}.

The inclusion of these mass-to-light ratios increases the number of
parameters to 8.  We tackle the problem of fitting these parameters in
two steps.  First, we fit the surface brightness profile to obtain the
scale radii $r_b$ and $R_d$ of the bulge and disk, and the luminosity
normalization of these components. The abundance and quality of the
surface brightness data is such that it would dominate the fit of the
scale radii in any case.  Also, the $\chi^2$ from this fitting step is
poor, which merely tells us that we have not managed to fit all the
details of M31's visual structure; in absolute terms, the size of the
discrepancies is actually fairly small.  Then, we incorporate the
dynamical information to fit the four remaining parameters (the bulge
and disk mass normalizations $M_b$ and $\Sigma_0$, and the two halo
parameters $r_h$ and $\delta_c$).  In contrast, the $\chi^2$ from 
this fitting step is actually meaningful, and we use it to constrain
the allowed region of parameter space.

We present the parameters for our best-fit solution in Table \ref{param},
and as we introduce below the set of observations that we use to constrain
the model, we also describe how well our best-fit model fares in
comparison.  We defer until \S~\ref{modelbehave} any discussion of
errors, correlations, and possible degeneracies between the
parameters, and the correspondence between the allowed terrain of
parameter space and physical properties of M31.

\begin{figure*}
\begin{center}
\leavevmode \epsfysize=9cm \epsfbox{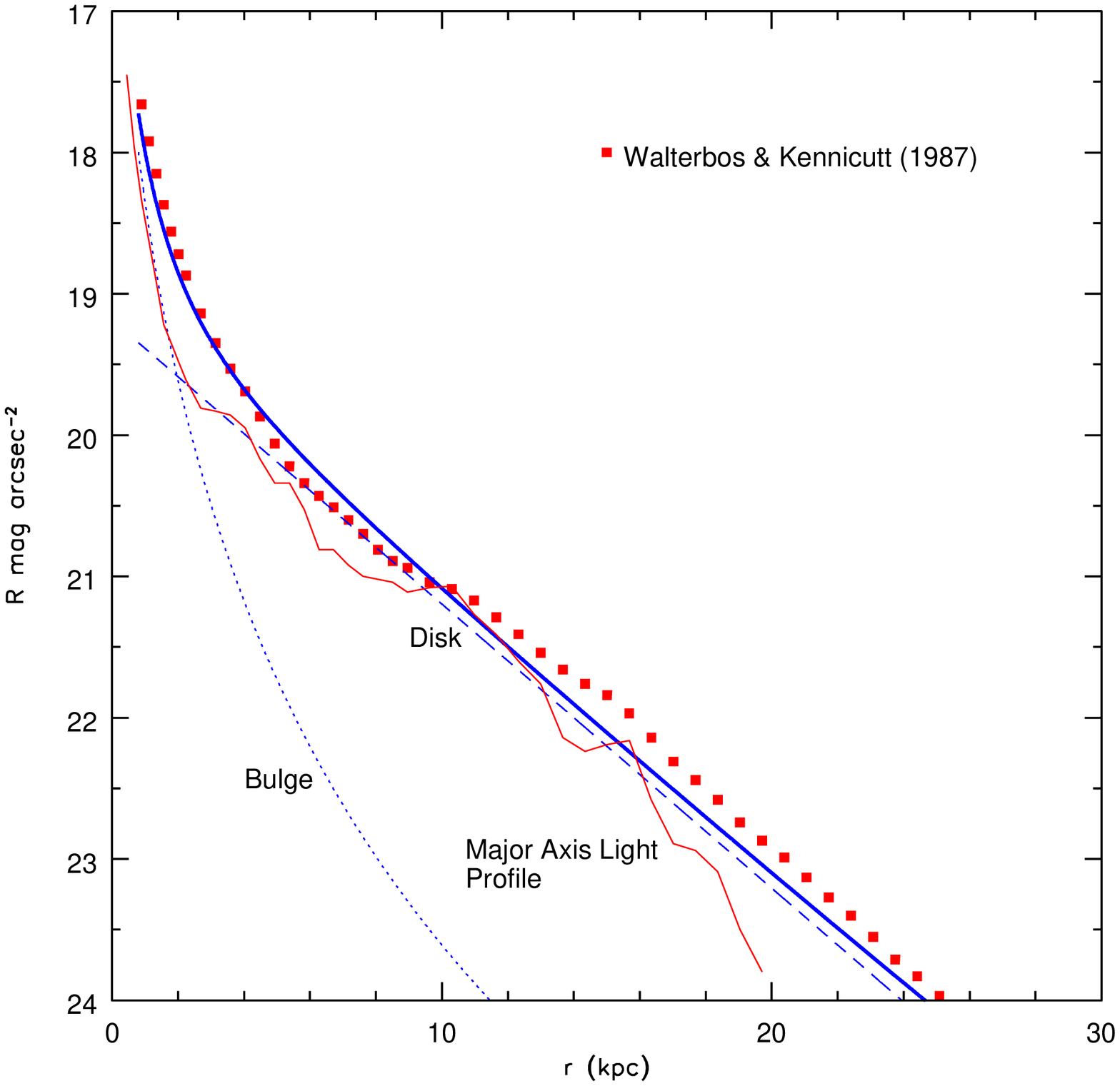}\epsfysize=9cm\epsfbox{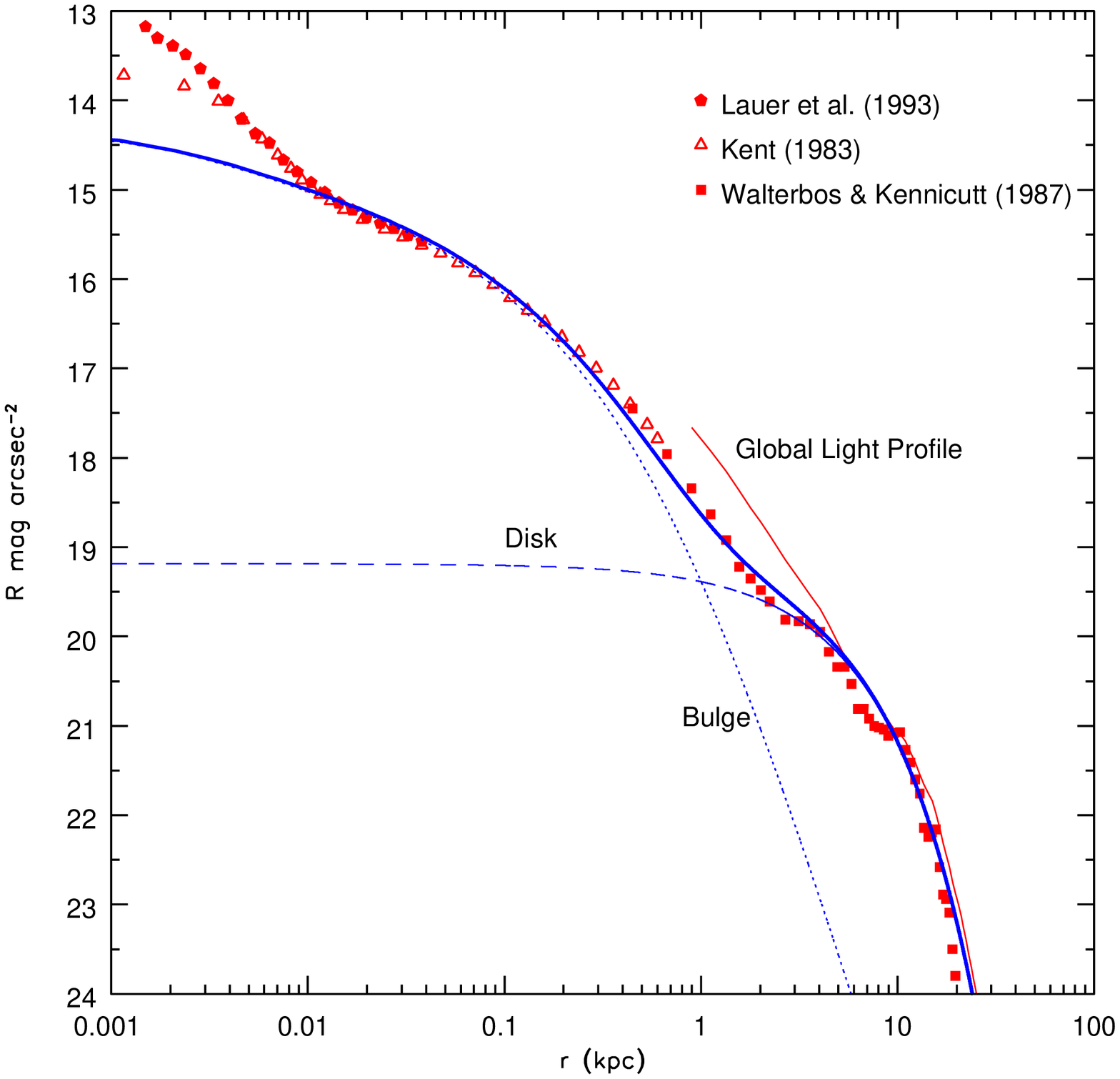}
\caption{
Left panel: The global surface brightness profile of M31 in log-linear
projection. The red squares represent the data of \protect\citet{wal87},
corrected as described in the text.  The blue lines show the results of our
best-fit model.  The dotted line is the bulge contribution, the dashed line the
disk and the thick solid blue line is the total.  For comparison purposes, we
juxtapose the observed major axis surface brightness data of
\protect\citet{wal87} from the right panel as the thin red line.  Right panel:
The symbols trace the composite, observed M31 major axis surface brightness
profile in log-log projection.  For comparison, we plot the observed global
light from the left panel as a thin red line.  As in the left panel, the blue
lines represent the major axis light profile for our best-fit model.
}
\label{surfbright}
\end{center}
\end{figure*}

\subsection{M31 Surface Brightness Data}\label{surfbprof}

Given its proximity, M31 is one of the few galaxies visible to the naked eyes.
Its existence was first documented by the Persian astronomer Abd al-Rahman Al-Sufi in
his treatise on stellar astronomy titled ``Kitab al-Kawatib al-Thabit al-Musawwar'' 
(Book on the Constellations of the Fixed Stars), published in AD 964, where
he both identified its position in the sky and summarized his observations.  
Though subject of some speculation and study over much of the ensuing millemium,
it was Hubble's epochal study \citep{hub29} establishing the true nature of M31 as 
an extragalactic stellar system (galaxy) that underlies the considerable effort
devoted to the study of M31 since the 1930s.  As a result, there is no 
dearth of high resolution photometric data on the system.  Here, we will use both the 
azimuthally-averaged (or ``global'') surface brightness profile and the major-axis 
profile to constrain our models.  The global profile has the advantage of minimizing 
effects from localized structural features, like the spiral arms, which can introduce 
bumps and wiggles in the light curve along any one direction.  In the left panel of
Figure~\ref{surfbright}, we plot the $R$-band global light profile of
\citet{wal87}.  We use the $R$-band profile because we expect it to be less
affected by dust extinction and stellar population variations than bluer bands.
For reasons discussed below, we find it necessary to shift the \citet{wal87} major
axis surface brightness profile by $-0.1$ mag, and since the major axis and the
present global light profiles are derived from the same photographic plates, we
apply the same shift to the global profile as well.  Walterbos \& Kennicutt
compute the mean surface brightness at radial coordinate $r$ by averaging the
light in an elliptical annulus formed by taking a face-on circular ring of radius
$r$ and thickness $\delta r$, rotating it to an inclination of 77$^\circ$,
projecting the resulting structure onto the plane of the sky, and aligning the
major axis of the resulting elliptical annulus with the major axis of M31. When
we compute the global surface brightness for our model, we follow the same
procedure.

The \citet{wal87} global light profile only extends down to just under 1 kpc 
along the major axis, and
therefore just begins to probe the region where the light is dominated by the
bulge.  On the other hand, there have been a number of studies of the light profile
along the major axis that probe all the way in to the very central nuclear
region.  We, therefore, also include the major-axis profile 
among our set of constraints.
We plot this profile in the right panel of Figure~\ref{surfbright}.
This combines the nuclear region {\it HST} data from \citet{lau93},
intermediate radii data from \citet{ken83}, and large radii data from \citet{wal87}.   
To convert Kent's $r$-band data and Lauer et al.'s $V$-band data into an $R$-band light profile, we use 
colours $V-R \approx 0.75$ \citep{ten94} and $r-R \approx 0.35$ \citep{jor94}.
As a check, we have tested these color transformations by
running a stellar burst model in the PEGASE.2 population synthesis code \citep{fioc97}.
For a burst time of 12 Gyr, we find agreement within $\pm$0.1 mag between our adopted offsets
between $V$, $r$ and $R$, as well as other measured colours of M31's bulge presented in Table \ref{M31color}.
Additionally, we found it necessary to translate the \citet{wal87} data by $-0.1$ mag in order
to bring it in line with the other two data sets. This level of fine-tuning is not surprising
given an uncertainty of $\sim 0.1$ mag in the colours as well as $\sim 0.1$ mag zero-point errors 
in the surface brightness measurements.

The major-axis light profile strongly indicates that the stellar nucleus
($r<0.01$ kpc) and the bulge are two separate components (see also
\citealp{kor99}).  As noted
previously, we do not attempt to model the nucleus because its mass is
too small to affect the masses of our galaxy components or orbits in
the halo region.  The major-axis profile also has a series of prominent
bumps not present in the global light curve, due to dust lanes, disk warps,
and spiral arms that intersect the major axis.  Azimuthal averaging
minimizes the presence of such features in the global profile.

We compute the model major-axis surface brightness profile as $\mu(r)=\mu_b(0.9r)+\mu_d(r)$, where
the disk 
and bulge surface brightness profiles in the $R$-band are
given by
\begin{align}
\mu_d(r) & =\left(\frac{M}{L_R}\right)_d^{-1} \Sigma_{0} e^{-r/R_{d}} {\mbox sec}(i), \label{disksurf}\\
\mu_b(r) & =2 \left(\frac{M}{L_R}\right)_b^{-1} \int_{r}^{\infty} \rho_b(x)\frac{x}{\sqrt{x^2-r^2}}\; dx. \label{bulgesurf}
\end{align}
Notice that in computing the surface brightness at projected distance $r$ along the major axis, we sum
the disk surface brightness at that projected distance with the bulge surface brightness at $0.9r$. We
do so in order to  compensate for the fact that in our model the light distribution is circularly
symmetric while the observed light distribution is ellipsoidal.  The factor $0.9$ comes from our
requiring that the area enclosed by an elliptical isophote at distance $r$ along the major axis is 
the same as the area of our equivalent circle.  This ``correction'' results in our circularly symmetric
bulge light profile being stretched outward.

When fitting the major axis and the azimuthally-averaged surface brightness
data,  we assign an uncertainty of $\pm
0.1$ mag arcsec$^{-2}$ to each data point.  The random errors are
probably a strong function of radius since they drop with surface
brightness, but there are significant systematic errors from
zero-point shifts and color correction terms in all parts of these
diagrams \citep{wal87}.  

Furthermore, there are many features that cannot be fit using our
simple model, and some of these are primarily features in the
luminosity (e.g., dust lanes) and not in the mass.
Our goal is simply to get a close approximation to the
surface brightness at all radii, not to follow every feature.
To this end, we use a standard $\chi^2$ fitting
routine to obtain the best-fit parameters, but ignore the actual value
of $\chi^2$ which indicates a poor fit (reduced $\chi^2$ is about 7).

The two panels of Figure~\ref{surfbright} show the global and major axis light
profiles for our best-fit mass model for M31.  Over most of the region of
interest, the model agrees with the data within 0.2 mag.  At first glance, the
systematic offset of $\sim 0.2$ mag at $r>10$ kpc between the model and the data
global surface brightness profiles may seem disconcerting.  This offset is due
to the fact that for $r>10$ kpc, the {\it observed} global light profile is
brighter than the major-axis profile by as much as 0.5 mag at $r=20$ kpc
\citep{wal87}.  To facilitate comparison, we show the trace of the observed
major-axis and the global light profiles in the left and right panels of
Figure~\ref{surfbright}, respectively.  If the disk was indeed an
infinitesimally thin, perfectly axisymmetric system and the bulge truly
spherically symmetric, the major-axis and the global light profiles of the disk
would be indistinguishable.  The observed differences in the light profiles are
due to the increased contribution to the global light profile by the bulge due
to its ellipsoidal shape, the warpage in the disk, and its finite thickness.
Since our mass model does not include these additional features, we do not
expect to be able to model the offset.  Instead, our best-fit model ``splits the
difference'' and settles in between the two.

\begin{table}
\caption{Observed M31 Bulge Colors}
\label{M31color}
{
\begin{tabular}{@{}ll}
\hline
Bulge Color  &   Reference  \\ 
\hline
$r-R \approx 0.35$ &  \citep{jor94} \\
$B-r \approx 1.28$ &  \citep{ken87} \\
$V-R \approx 0.75$ & \citep{ten94} \\
$B-R \approx 1.7 $ &  \citep{wal87} \\
$B-V \approx 0.97$ &  \citep{wal87, ten94} \\
$V-K \approx 3.38$ &  \citep{prit77} \\
$V-I \approx 1.34$ &  \citep{lau93} \\
$g-r \approx 0.55$ &  \citep{hoe80} \\
$v-g \approx 0.78$ &  \citep{hoe80} \\
\hline
\multicolumn{2}{l}{$^{\ast}$ All colors have an uncertainty of $\pm$0.1}
\end{tabular}
}
\end{table}

\subsection{Disk Rotation Curve}\label{rotprof}

\begin{figure*}
\begin{center}
\leavevmode \epsfysize=9cm \epsfbox{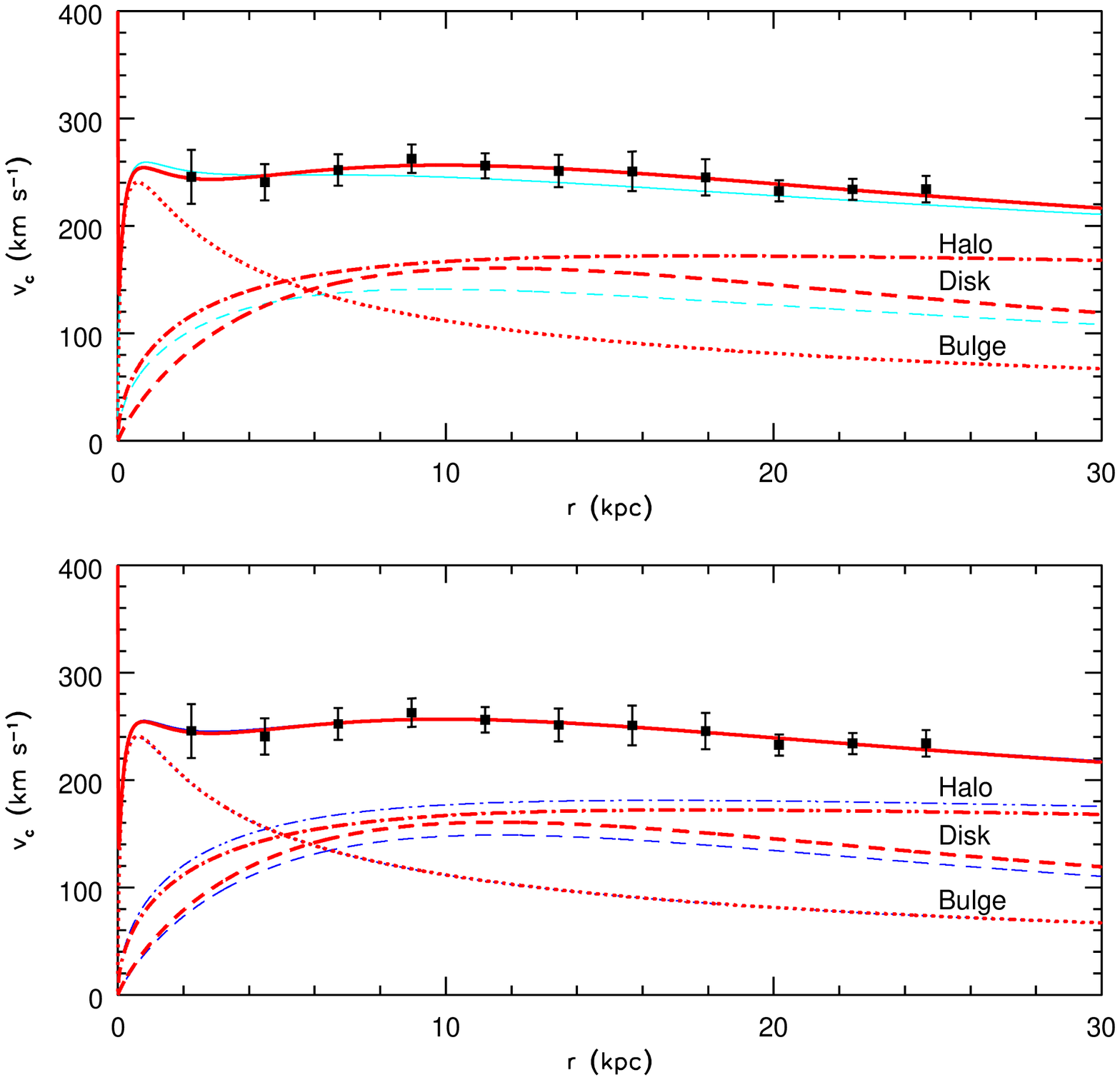}\epsfysize=9cm\epsfbox{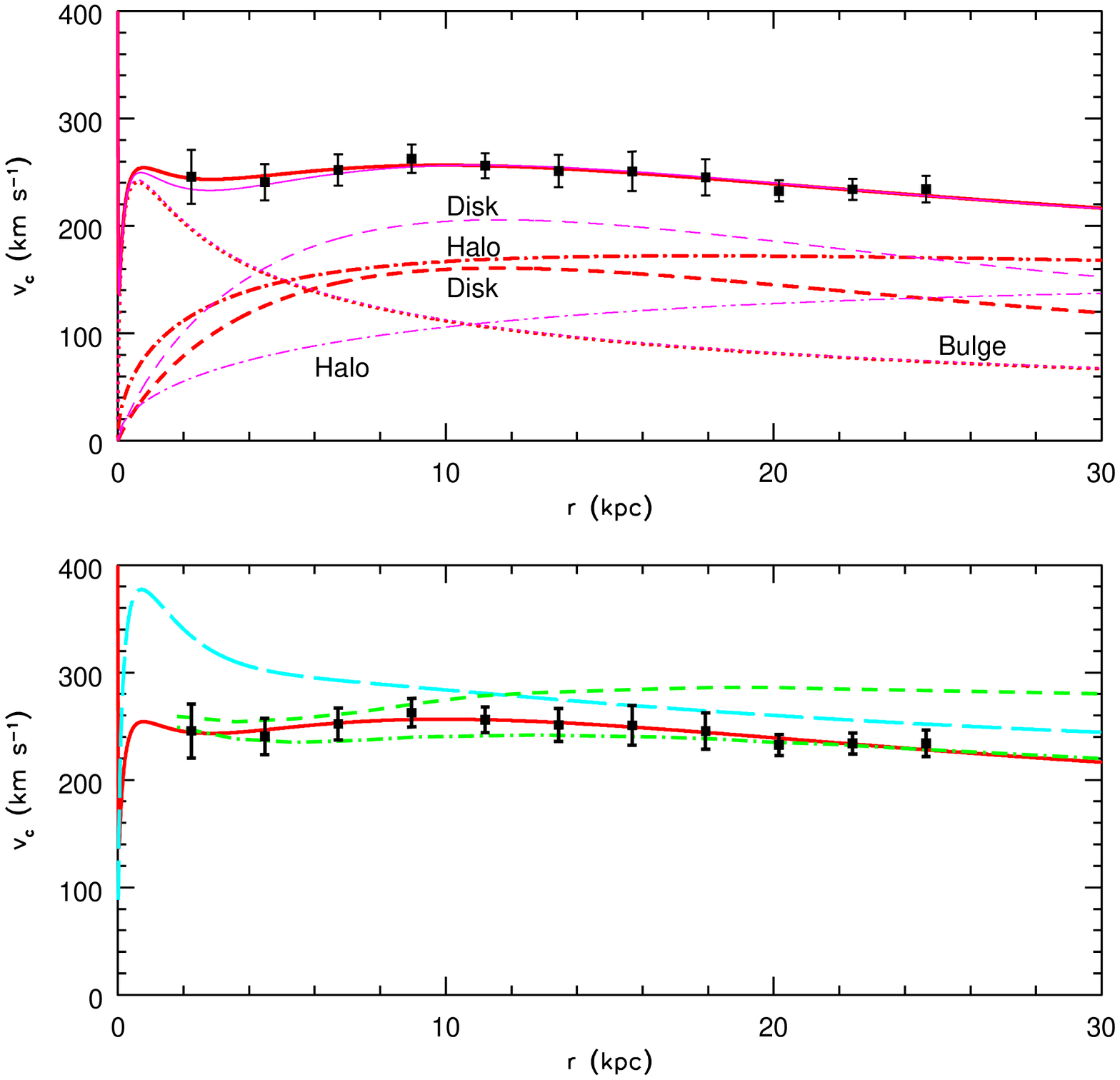}
\caption{
The top left panel shows our best-fit axisymmetric (thick red lines)  and spherical 
disk (thin cyan lines) model rotation curves, in comparison to the M31 disk rotation 
profile.  The solid curves show the total model rotation curves. The short dashed 
lines show the contribution to the total from the disk, the dotted 
lines, from  the bulge, and the dot-short dashed lines, from the halo. 
The spike in the curves at $r=0$ is due to the presence of the BH.
The bottom left panel compares the best-fit axisymmetric and constrained (M/L=3.3)
disk models.  As in the top panel, the thick red lines correspond
to the best-fit axisymmetric model.  The thin blue lines show the results for the M/L=3.3
model.  The line types are the same as in the top panel.
The top right panel compares the best-fit axisymmetric (thick red lines) and 
the maximum disk (thin magenta lines) models.  Line types are the same as discussed above.
The bottom right panel compares our best-fit model rotation curve (thick red lines)
against those of recently used models: the \protect\citet{bek01} model used by Font et al. (cyan 
long-dashed line), as well as the high (green short dashed line) and the low (green dot-dashed
line) \protect\citet{klyp02} models used by Ibata et al.
}
\label{rotplot}
\end{center}
\end{figure*}

Figure~\ref{rotplot} shows the rotation profile of the M31 disk.  As
with the photometric data, there is no dearth of optical and radio
observations of the rotation data for the M31 disk.  We use the
smoothed, composite rotation curve of \citet{wid03}, which was based
on the HII emission region observations of \citet{ken89} 
and the HI measurements of \citet{brau91}.

The two left panels and the top right panel of  Figure~\ref{rotplot} 
also show the total disk rotation curve of our best-fit model (thick solid curve),
along with  the contributions from the bulge, disk and halo components (thick
dotted, thick dashed and thick dot-dashed curves, respectively).  Our model 
results are in excellent agreement with the observations 
and we remind the reader that the corresponding model parameters 
are listed in Table~\ref{param}. 

In the top left panel, we  also show the results of a model that we have dubbed 
the ``spherical disk'' model.   The halo and the bulge components in this
model  are exactly the same as in the best-fit axisymmetric model.  In fact,
even the total radial mass profile for the disk component, $M_{d}(<r)$ is the 
same in the two cases.  In the spherical disk case, however, this mass is 
assumed to be distributed in a spherically symmetric fashion, as opposed to 
planar and axisymmetric.  As Figure~\ref{rotplot} illustrates, the rotation 
profiles for the two models (and hence, the potential, at least in the 
plane of the disk of the axisymmetric case)  are similar with minor 
differences being solely due to the different geometries of the disk
mass distribution.  Since orbit calculations in purely spherical potentials are 
much faster to compute, we use the spherical disk model to narrow the space of 
potentially acceptable satellite orbits and then, use the axisymmetric model to 
explore this limited space further.   This latter step is necessary because as we
demonstrate in \S\ref{others}, the nature of the disk potential, whether
axisymmetric or spherically symmetric, impacts the satellite orbits in
non-trivial ways.

In the bottom left panel, we juxtapose our best-fit axisymmetric model against 
the results for a model where the disk mass-to-light ratio has been constrained
to be $M/L_R=3.3$, which is lower than the mass-to-light ratio of the disk in the 
best-fit case.  In the top right panel, we juxtapose our best fit-axisymmetric 
model against the results for a model, which we will refer to as the ``maximum
disk'' model, where the disk mass-to-light ratio has been 
constrained to be $M/L_R=6.3$, a value that is much higher than the $M/L$ of 
the disk in the best-fit case.  The parameters for the two constrained models  are also 
given in Table~\ref{param}, and reasons for considering these two additional models 
are discussed in \S\ref{solnature}.  Here, we merely note that total rotation 
profiles for the formal best-fit model, the constrained-$M/L$ model and the 
maximum disk model are virtually indistinguishable.
Fixing the disk mass-to-light ratio forces the disk mass 
higher or lower relative to the best-fit model. The corresponding change in the disk 
contribution to the total rotation profile is entirely compensated by an opposite 
change in the halo contribution.   This playoff between the disk and the halo is 
an example of the degeneracy that, in the absence of any additional constraints, 
is a fundamental source of uncertainty affecting all efforts to construct mass
models of disk galaxies.  We discuss the nature of,  as well as 
our efforts to break this degeneracy  in \S\ref{solnature}.

\subsection{Bulge Velocity Dispersion}\label{bveldisprof}

\begin{figure}
\begin{center}
\includegraphics[width=84mm]{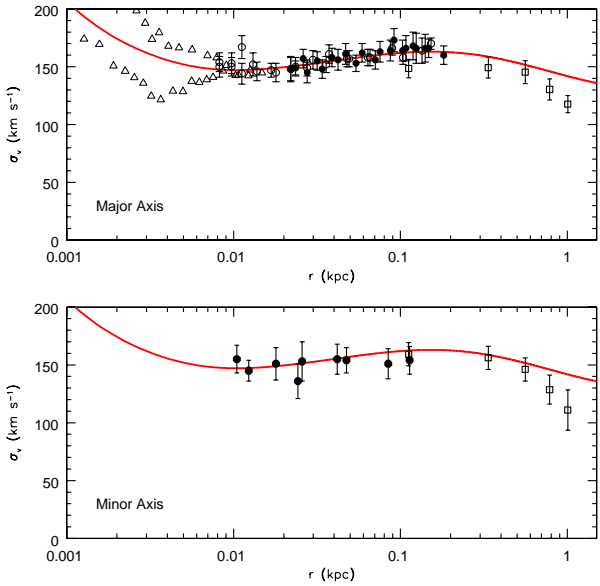}
\caption{
The upper and lower panels show the measured bulge velocity dispersion
along the bulge major and minor axis, respectively.  The solid curve
represents our best-fit model.  The filled circles represent data
of \protect\citet{kor88} 
along PA=55$^\circ$ (upper panel) and 
along PA=160$^\circ$ (lower panel).  
Also plotted are the major and minor axis velocity dispersion
data of \protect\citet{mce83} as smoothed by \protect\citet{wid03} (open squares), 
the major axis data of \protect\citet{kor99} (open triangles), and the major
axis data of \protect\citet{kor88} along PA=38$^\circ$ (open circles).  
The upturn in the curves at $\sim 0.01$~kpc is
due to the influence of the central BH.
}
\label{veldisplot}
\end{center}
\end{figure}

In Figure~\ref{veldisplot}, we show the projected bulge velocity
dispersion profile along the bulge major (PA=45$^\circ$---55$^\circ$)
and minor (PA=135$^\circ$---160$^\circ$) axes.  The former is plotted in 
the upper panel while the latter is shown in the lower panel.  
The filled data points (circles) with error bars show the observed
dispersion that we are using in our fitting procedure.  These correspond
to the measurements by \citet{kor88} along PA=55$^\circ$ (major)
and PA=160$^\circ$ (minor).  We only consider measurements at
$r\gta 0.02\kpc$ to avoid complications due to the BH.

For comparative purposes only, we also include on the plot data points 
from \citet{kor88} taken along PA=38$^\circ$ (open circles) as well as
the data from \citet{kor99} (open triangles).  The latter extends the
velocity dispersion profiles well into the nucleus of M31 and the 
trend in the measurements towards smaller radii reflects the increasing
gravity due to the BH.  Focusing on the PA=38$^\circ$ data from
\citet{kor88} (open circles), we note that this latter set is indistinguishable 
from that for PA=55$^\circ$ in the region of overlap, suggesting that small 
variations in the position angles along which the measurements are made or 
the fact that these measurements may not be mapping out the velocity 
dispersion profile precisely along the ``true'' bulge major axis does not 
appear to introduce any significant biases.   

For completeness, we also show the data points (open squares) corresponding to 
the smoothed profile of \citet{wid03}, which itself is based on data from 
\citet{mce83}.  We only  show the data only out to 1 kpc, since beyond this point 
we are concerned about the dynamical effect of bulge rotation and/or
contamination from the disk.  There is one aspect of this data set that is 
slightly disconcerting.  In the region of overlap along the major axis, 
the \citet{kor88} data implies $\sim$15\% higher velocity dispersion than 
the \citet{mce83} observations.  We do not understand the origin of 
this disagreement.

The solid curves in the two panels show the velocity dispersion for
our best-fit mass model. The results match fairly well to the data points 
actually used to constrain our model.  In keeping with the
observations, the model profiles decline towards increasing radii
beyond $r=0.2$ kpc though the decline is not as steep as indicated by 
\citet{mce83}.  Moreover, although the model was not
constrained to fit 
the \citet{kor99} data,  it
nonetheless reproduces the gentle decline towards the center inward of
0.1 kpc, the trough at $\sim$0.01 kpc, and the subsequent rise into
the center.

In computing the model velocity dispersions, we have made three
additional simplifications, in keeping with our basic assumption of a
spherically symmetric bulge.  These additional assumptions are: (a)
The bulge velocity dispersion is isotropic; (b) the disk potential in
the central 1.2 kpc can be approximated by that of the spherical disk;
and (c) the bulge rotation can be neglected.  Then the true
(unprojected) bulge velocity dispersion is
\begin{equation}
\sigma_{r}^{2}=\frac{1}{\rho_{b}(r)}\int_{r}^{\infty}{\rho_{b}(r')\frac{\partial \Phi_{tot}}{\partial r'} dr'}.
\end{equation}
The observations, however, yield the luminosity-weighted {\it projected}
velocity dispersion.  The equivalent model profile is \citep{sim79,
ken89}
\begin{equation}
\sigma_{p}^{2}(r)=\frac{1}{\mu_b(r)}\left(\frac{M}{L_R}\right)_b^{-1}
\int_{r}^{\infty}{\rho_{b}(r')\sigma_{r}^{2}(r'){\frac{r'}{\sqrt{{r'}^{2}-r^2}}}}\;dr',
\end{equation}
where $\mu_b(r)$ is the bulge surface brightness profile given in
\eqref{bulgesurf}. Since the projected velocity dispersion is
luminosity-weighted and since our model isophotes are circular whereas
the actual bulge isophotes are elliptical, 
it would seem that 
we perhaps ought to rescale
our profiles so that along the major axis, $\sigma_p(r)$ is stretched
out by $\sim$10\% and along the minor axis, $\sigma_p(r)$ compressed
by the same amount.  However, we omit this step, since $\sigma_p(r)$
is so flat that this rescaling has a negligible effect.  

After comparing our results to those obtained with more sophisticated
assumptions about the bulge geometry and velocity distribution, we find that the
three assumptions listed above, especially assumptions (b) and (c),
do not appear to affect 
our results significantly.
We base this assertion on the discussion in
\citet{sim79} and \citet{ken89}.  The former, for example, demonstrate that
impact of rotation on the determination of $\sigma_{p}^{2}(r)$ is negligible for
$r<0.3$ kpc.  Between 0.3 and 1.0 kpc, the bulge rotation profile rises from
$V_{c,b}\sim 0.3\sigma_p$ to $0.5\sigma_p$.  It might be thought that
this increases the support from rotation and induces the slight
discrepancy between our models and the data at~$\sim 1 \kpc$.  
As reasonable as this sounds, the argument is not supported by the results of
\citet{ken89}. In constructing his mass model, Kent assumed, like us,
an isotropic bulge velocity dispersion but, unlike us, he also 
used an oblate spheroidal model for the mass distribution 
of the M31 bulge, explicitly took
into account the axisymmetric nature 
of the disk potential, and allowed for bulge rotation.  
Kent's velocity dispersion profiles are, however, similar to
ours both in shape and magnitude.
In particular, the Kent model profiles also do not
fall off as rapidly as the observed profile.  To the extent that we have an
excellent overall agreement with the more rigorous calculations of
\citet{ken89}, we feel justified in having made our simplifying assumptions. 
A clue to the origin of the steeper than predicted
decline in $\sigma_{p}^{2}(r)$ may lie in the different rates at which the major
and minor axis profiles fall off.  Both \citet{sim79} and \citet{ken89} note
these differences may be indicative of velocity anisotropy.

We note that our neglect of bulge rotation is primarily motivated by consistency
with our assumption of a spherical bulge, and by its small level
relative to the bulge velocity dispersion.  Additionally,
\citet{mce83} noted several asymmetries in the rotation curves on the
two sides of the putative axis of rotation as well as along various
different position angles.  This suggests the determination of a mean
bulge rotation velocity is not straightforward, which also encourages us 
to neglect it.

\begin{figure}
\begin{center}
\includegraphics[width=84mm]{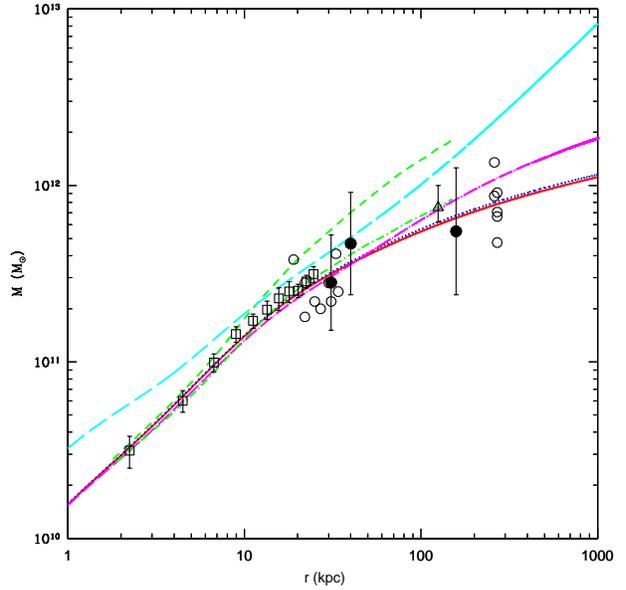}
\caption{
A comparison of the mass profiles for different models: our formal best-fit
model (solid red line), our best-fit models for ($M/L_{R}$)$_{d}$=3.3 (dotted blue line) 
and the  maximum disk (dot-long dashed magenta line), the two axisymmetric models of 
Ibata et~al.\ based on the mass models of \protect\citet{klyp02} (the dashed and dot-dashed
green lines), as well as the bulge-disk-halo model of \protect\citet{bek01} used by
Font et~al.\ (long dashed cyan line).
The solid circles represent mass estimates
made by \protect\citet{eva00} and \protect\citet{ew00} that we used as constraints on
our mass models.  The first two solid circles are mass estimates from
planetary nebula and globular cluster tracers, respectively, from
\protect\citet{ew00}.  The last solid circle is a mass estimate based on M31
satellite galaxies from \protect\citet{eva00}.  The open squares represent the
mass estimate based on the observed rotation profile computed using the
spherical approximation.  The open triangle represents an estimate of
the mass within 125~kpc derived by \protect\citet{iba04} while the open
circles represent mass estimates from many sources which are
summarized in Table~6 of \protect\citet{ew00}.  None of the mass estimates
shown as open symbols were used as constraints on our mass model.
}
\label{massplot}
\end{center}
\end{figure}

\subsection{Total Mass Estimates from the Intermediate  and Outer Halo Regions}\label{massprof}

Finally, we consider the constraints on the total mass of the system from
dynamical studies of globular clusters, planetary nebulae and M31 satellites.  A
number of researchers have used these tracers to estimate the total mass profile
of M31 in the intermediate and the outer regions of the halo.  However, few of
these estimates contain a well-motivated error bar, which we need for our
fitting procedure; many authors simply propagate the observational errors
through their mass estimator and ignore the larger statistical errors
altogether.  For the present purposes, we thus use the results of \citet{ew00}
and \citet{eva00}, who estimate the errors using Bayesian statistics.  Using the
planetary nebulae and globular cluster data, \citet{ew00} estimate that
$M(<31\;{\rm kpc})=(2.8^{+2.4}_{-1.2})\times 10^{11}\; \msun$ and $M(<40\;{\rm kpc})
=(4.7^{+3.4}_{-2.3})\times 10^{11}\; \msun$, respectively.  Of these two, \citet{ew00}
argue that the mass estimate based on globular clusters is more robust because
these tracers are more uniformly distributed while the planetary nebulae tend to
be concentrated in two regions near the optical disk.  To estimate the mass
further out in the halo, \citet{eva00} analyzed the data for the satellite galaxies of M31, 
including the spiral galaxy
M33.  We use their results to estimate the mass at 158
kpc, which is the radius of the median object (And II) in the sample of 13 M31
satellites, {\it excluding} Pegasus and IC 1613.  These two satellites appear to
be outside M31's virial radius and thus are not appropriate to include in an
analysis that assumes statistical equilibrium.  We obtain the mass estimate by
averaging the results of their two velocity distributions, and making a small
correction from their stated total mass to $M(<158\;{\rm kpc})$ using their mass
profile.  This gives $M(<158\;{\rm kpc})=5.5^{+7.1}_{-3.1}\times 10^{11}\;
\msun$.  For each of the three halo mass estimates, we convert the given
confidence intervals to symmetrical error bars in $\log(M)$ for use in our
$\chi^2$ fit.

In Figure~\ref{massplot}, we show the above three mass estimates as
filled circles.  These are the values that we use to constrain
our mass models.  For illustrative purposes, however, we also plot a
number of other mass estimates.  The open squares extending from
$\sim$2 to $\sim$30 kpc trace the mass estimates based on the
rotational curve data, computed using the spherical approximation
$V_c^2 = GM/r$.  
The open triangle at 125 kpc designates the mass estimate derived by
\citet{iba04} from their preliminary analyses of the dynamics of the
giant southern stream.  We shall consider this further in Paper II.
Finally, the open circles represent
mass estimates by a number of different researchers, taken from
Table 6 of \citet{ew00}.

Figure~\ref{massplot} also shows the results of our best-fit model 
(solid red curve)
and two variants, 
one with the disk mass-to-light ratio fixed to ($M/L_{R}$)$_{d}$=3.3 
(dotted blue line) and the maximum disk model (dot-long dashed magenta line).
All three curves are in excellent agreement with not only the three data
points used as constraints, but also in good agreement with the entire
slew of mass estimates shown in the figure.  
The curves, however, do lie systematically below the mass estimates computed
from the observed rotation profile using the spherical approximation.    
Since the rotation curve associated with these models are in excellent agreement 
with the observed profile, this comparison shows clearly the effect of neglecting disk 
flattening.  The virial mass ($M_{200}$) corresponding to the three axisymmetric
models spans the range $6.8$---9.4$\times 10^{11}\msun$ (see Table~\ref{param}).

\section{Properties of Allowed Models}\label{modelbehave}

The parameters corresponding to our formal best-fit mass model for M31 are listed 
in Table \ref{param}.   The table also shows parameters corresponding to two alternatives
to our formal best-fit model, one where the disk mass-to-light is fixed to 
($M/L_{R}$)$_{d}$=3.3 and another where the disk mass is fixed to the maximum allowed 
value.   We discuss below the reasons for considering these alternative solutions.

\begin{figure}
\begin{center}
\includegraphics[width=84mm]{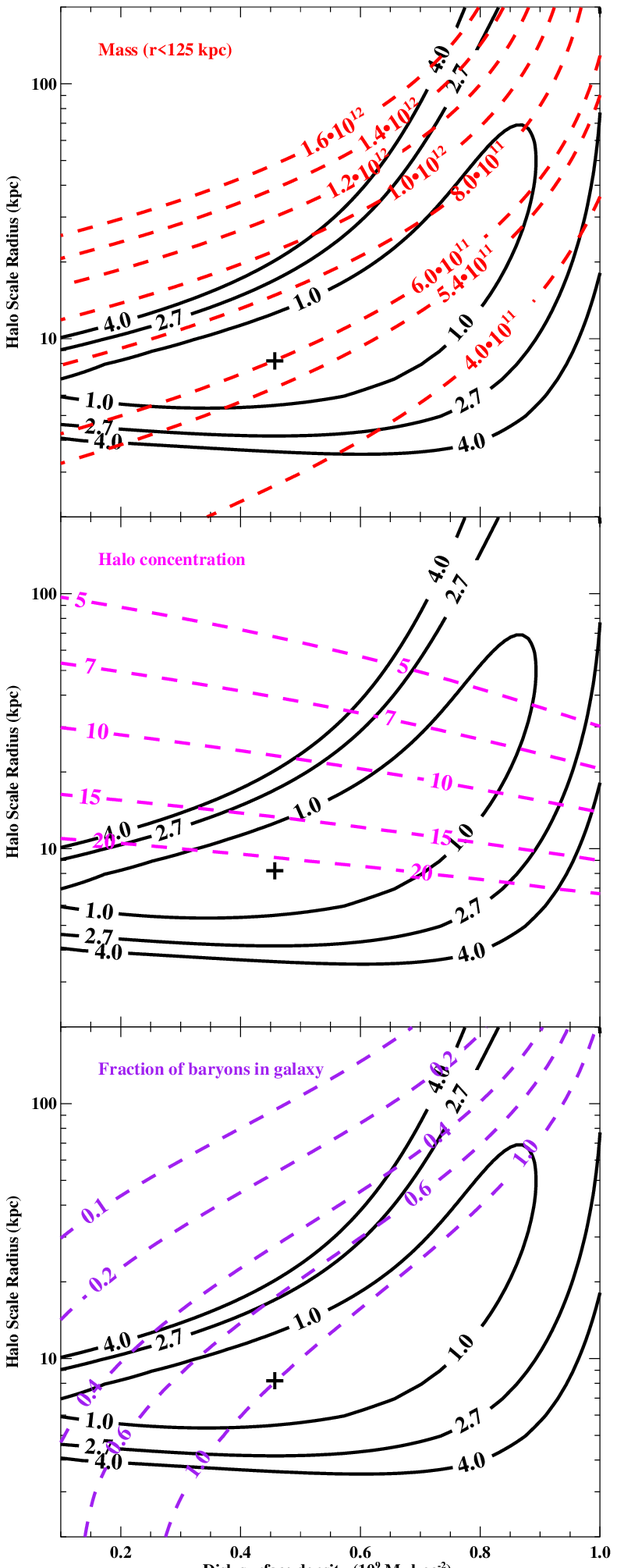}
\caption{The cross in the above three panels marks the location of the formal 
best-fit solution (see Table~\ref{param}), and the solid black curves show
the corresponding $\Delta \chi^{2}=$1.0, 2.7, 4.0 contours. 
The upper panel superposes
the contours corresponding to constant total mass within 125~kpc (red dashed lines);
the middle panel shows lines of constant halo concentration $C_{200}$ 
(magenta dashed lines); and the lower panel shows lines of constant baryon fraction 
(purple dashed lines) as defined in the text.
}
\label{contour}
\end{center}
\end{figure}

\subsection{Allowed Region of Parameter Space}\label{solnature}

In general, attempts to model the mass distribution in galaxies are hampered by
degeneracies between the various parameters. In our specific case, we had
expected to encounter a linked bulge-disk-halo degeneracy where the halo and
the disk played off against each other to match the M31 rotation profile, while
the disk and the bulge did the same with respect to the inner region dynamics.
In fact, our combined use of photometric and kinematic data sets greatly
reduces the available degrees of freedom.

Of the 8 parameters, two --- the disk and bulge mass-to-light ratios --- are not
really independent variables. They appear in Equations~\eqref{disksurf} and
\eqref{bulgesurf} as multiplicative factors relating disk and bulge mass
profiles to the M31 light profiles. Consequently, once the disk parameters
($\Sigma_0$, $R_d$) are determined, the M31 surface brightness profiles fix the
disk mass-to-light ratio to high accuracy. 
The same is true of the bulge. 
The light profiles also
determine the bulge and disk scale radii fairly precisely to 0.61$\pm$0.10 and 5.40$\pm$0.13 kpc,
respectively.  We therefore fix $r_b$ and $R_d$ to these values after doing the
surface brightness fit, removing these particular degrees of freedom entirely.
To estimate the errors, we have adopted the
approach advocated by \citet{numrec} and rescaled the error estimate associated with 
individual data points such that the effective reduced $\chi^2$ is unity.  
As noted in \S\ref{surfbprof}, our fit to the M31's 
light profile is poor in a formal sense using the original photometric errors (the corresponding reduced $\chi^2$ is 7) 
because we do not attempt to fit the various visual features present therein. 
By increasing
the error bars, we are effectively rendering the physical variations in the light curves
due to dust lanes, disk warps and spiral arms statistically insignificant.
We then use this rescaled 
$\chi^2$ to approximate the uncertainties in the bulge and disk scale radii 
estimates.

\begin{table*}
\caption{M31 Mass Model Parameters for Best-fit, disk with ($M/L_{R}$)$_{d}$=3.3, and Maximum disk cases}
\label{param}
{
\begin{tabular}{@{}lllr@{.}lr@{.}lr@{.}l}
\hline
Parameter & Symbol & & \multicolumn{2}{l}{Best-fit Model} & \multicolumn{2}{l}{($M/L_{R}$)$_{d}$=3.3} & \multicolumn{2}{l}{Maximum Disk}\\
\hline
\vspace{3mm}
Black Hole Mass               & $M_{\bullet}$              & ($10^{7}\ \; \msun$)        & 5&6           & 5&6 &      5&6\\
Total Bulge Mass              & $M_b$                 & ($10^{10}\; \msun$)         & 3&3           & 3&2 &      3&3\\
Total Disk Mass               & $M_d$                 & ($10^{10}\; \msun$)         & 8&4     	    & 7&2 &      13&7\\
Total Mass inside 125 kpc     & $M(<125\;{\rm kpc})$  & ($10^{11}\; \msun$)         & 5&6           & 6&2 &      7&3\\
Virial Mass                   & $M_{200}^\ast$        & ($10^{11}\; \msun$)         & 6&8           & 7&1 &      9&4\\
\vspace{3mm}
			      & $M_{100}^\ast$        & ($10^{11}\; \msun$)         & 7&5	    & 7&7 &      10&8\\
Bulge Scale Radius            & $R_b$                 & (kpc)                       & 0&61          & 0&61 &     0&61\\
Disk Scale Radius             & $R_d$                 & (kpc)                       & 5&4           & 5&4  &     5&4\\
Halo Scale Radius             & $r_h$                 & (kpc)                       & 8&18          & 7&63 &     28&73\\
Virial Radius                 & $R_{200}^\ast$        & (kpc)                       & 180&0         & 182&3 &    200&3\\
\vspace{3mm}
                              & $R_{100}^\ast$        & (kpc)                       & 231&8         & 234&8 &    262&4\\
Halo Density Parameter        & $\delta_c$            & ($10^4$)                    & 27&0          & 34&4 &     1&54\\
\vspace{3mm}
Halo Concentration Parameter  &$C_{200}\equiv R_{200}/r_h$    &                     & 22&0          & 23&9 &     7&0\\
Bulge M/L                     & $\left(M/L_R\right)_b^{\ast\ast}$&                  & 3&9           & 3&9 &      4&0\\
\vspace{3mm}
Disk M/L                      & $\left(M/L_R\right)_d^{\ast\ast}$&                  & 3&9           & 3&3 &      6&3\\
Disk Central Surface Density  & $\Sigma_0$            & ($10^{8}\; \msun/$kpc$^2$)  & 4&6           & 3&9 &      7&5\\
\vspace{3mm}
Maximum Rotation Velocity     & $V_{c,max}$           & km/s                        & 256&5         & 255&9 &    256&5\\
Fraction of ``galactic'' baryons & \multicolumn{2}{l}{$\Omega_m(M_b + M_d)/(\Omega_b M_{200})$}     & 0&99          & 0&86 &     1&0\\
\hline
\multicolumn{9}{l}{$^{\ast\phantom{\ast}}$ We define $M_\Delta$ as the mass enclosed with the sphere of radius $R_\Delta$ such that the mean density inside is $\Delta\rho_c$, where $\Delta=100$ or $200$} \\
\multicolumn{9}{l}{$\phantom{^{\ast\ast}}$ and  $\rho_{c}=277.72h^2\;\msun/\kpc^2$ is the present-day critical density.}\\
\multicolumn{9}{l}{${^{\ast\ast}}$ The quoted M/L ratios are based on luminosities that have {\it not} been corrected for internal or foreground extinction.}  \\
\end{tabular}
}
\end{table*}

Additionally, we find that the two halo parameters $\delta_c$ and
$r_h$ are coupled so strongly that if all other parameters are already
fixed, setting one effectively sets the other. The main reason for
this is the small error bars on the rotation curve points, which put
tight constraints on the halo contribution there. The constraints on
the outer halo force are not as strong, as can be seen in
Figure~\ref{massplot}.

Moreover, the solutions with the lowest $\chi^2$ lie on a thin plane
corresponding to a specific set of bulge masses $M_b$.  The values of this
parameter are remarkably robust in that forcing large changes in the values of
the other parameters result in changes of $<10$\%. This was previously noted
by \citet{wid03}, who described the neighbourhood of their best-fit solution as a
``trough [in the $M_d$--$M_b$ plane] running parallel to the $M_d$ axis.''

This then leaves just two parameters, $\Sigma_0$ and one of the two
halo parameters, to span most of the available parameter space. In
Figure~\ref{contour}, we show the location of 
our formal best-fit solution and the corresponding
$\chi^2$ contours in the
$\Sigma_0$--$r_h$ plane. Here we show contour levels of $\Delta\chi^2
\equiv \chi^2 -\chi^2_{\rm min} = 1.0$, 2.7, and 4.0, corresponding to
68\%, 90\% and 95\% confidence intervals in the case of one parameter,
and 39\%, 74\%, and 86\% in the case of two parameters
\citep{numrec}.  (Initially we regard both parameters as free, but
later we will assume a value for the disk $M/L$, resulting in just
one free parameter.)
This plot shows that there is a huge area of the
plane that produces acceptable fits, although the transverse width of
this allowed region in the remaining dimensions of parameter space is
rather small.  The importance of the formal best-fit solution is thus
diluted by this large allowed volume, and it is worth seeing whether we
can put any more constraints on the family of solutions as a whole.

In the upper panel, we superpose on
the contour map lines of constant total mass within 125 kpc. 
The shape of the contours shows that the main degeneracy is the
tradeoff between the disk and halo components; this is not very
surprising, as this tradeoff is a long-standing problem in fitting
galaxy rotation curves \citep{dutton05}. 
The allowed region extends from ``maximum
disk'' models at the upper right, to models where the disk is a minor
contributor to the rotation curve at the lower left.
The extension in the other direction, across the mass contours,
is due to the large uncertainty in the outer halo mass, which allows for
a large tradeoff between $r_h$ and $\delta_c$.

The best-fit solution lies on the line for 
$M(<125\;{\rm kpc})=6.1\times 10^{11}\; \msun$. 
As already noted, \citet{iba04} have carried out a preliminary analysis of 
the dynamics of the giant southern stream, a separate constraint to those
examined here, and their results require that 
$M(<125\;{\rm kpc}) > 5.4\times 10^{11}\;\msun$. This
constraint reduces the region of allowed parameter space by excluding the lower
right corner region of the $\Delta\chi^2 = 4.0$ confidence region, along with the entire
lower right quarter of the $\Sigma_0$--$r_h$ plane. 

In the middle panel of Figure~\ref{contour}, 
we superpose lines of constant halo concentration
$C_{200}$ on the contour map. (Our notation emphasizes we use the radius
enclosing a mean density of 200 times the critical density to define the virial
radius and concentration, in contrast to some other authors.)  Numerical
simulation studies suggest that in the absence of baryons, the concentration
parameters of $\sim 10^{12}\;\msun$ halos are in the range $C_{200}\approx 7$--$14$
\citep{bul01,wechsler02,dolag04}.  If anything, the inclusion of baryons ought
to make the halos more concentrated due to gravitational compression of the 
inner regions of the halo, following the expected cooling and pooling of the 
baryons at the centre of the halos \citep{blumenthal86,klyp02}.  On the other 
hand, heating of the dark matter by gas clumps during galaxy formation 
\citep{mo04} and by a bar in the central disk \citep{weinberg02} may partially 
counteract this effect.  For these reasons, we use $C_{200} >7$ as a lower limit on 
the effective dark halo concentration.  This condition excludes the upper third of 
the $\Sigma_0$--$r_h$ plane, including some of the $\Delta\chi^2 = 4.0$
confidence region.

\begin{figure}
\begin{center}
\includegraphics[width=84mm]{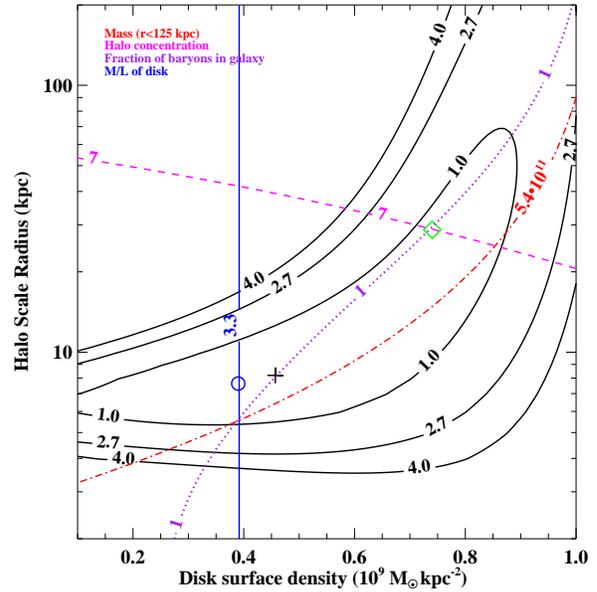}
\caption{
The above plot again shows our
formal best-fit model (cross) and the
$\Delta \chi^{2}=$1.0, 2.7, 4.0 contours (black solid lines).
However, we have also combined, for clarity,
the three critical demarcations from the panels in Figure~\ref{contour} (see
text).  We assert that the allowed region for physically plausible solutions is
the section of the $\Delta \chi^{2}<$4.0 region that is bounded at the top by
$C_{200}=7$ (magenta dashed curve) and to the right by $f_{\rm gal} =1$ (purple
dotted line).  Additional lines on the plot correspond to $M(<125\;{\rm
kpc})=5.4 \times 10^{11}\;\msun$ (red dot-dashed curve) and disk mass-to-light
ratio (uncorrected for extinction) of $(M/L_R)_d=3.3$ (solid blue vertical
line).  Our constrained, best-fit, physically plausible solution
is shown as an open circle on the $(M/L_R)_d=3.3$ line.
Our constrained, maximum-disk solution is shown by the green diamond.
}
\label{constrain}
\end{center}
\end{figure}

Finally, the lower panel superposes the contours of the fraction
of baryons within the virial radius $R_{200}$ that are ``galactic'',
i.e., within the bulge or the disk component: 
$f_{\rm gal}=\Omega_m(M_d+M_b)/(\Omega_b M_{200})$. In computing this 
quantity, we assume that the bulge and the disk are purely baryonic. 
This is a long-debated assumption even in the case of the 
Milky Way, but the disk and bulge are probably not {\em dominated} by 
dark matter so that the error from this assumption is probably small.  
Since $f_{\rm gal}$ must be less than unity, much of the lower right half 
of the $\Sigma_0$--$r_h$ plane is excluded.  For $\Sigma_0 > 4.5\times 10^{8}
\; \msun/{\rm kpc}^2$, this condition is stricter than that of
$M(<125\;{\rm kpc})$. 
Figure~\ref{constrain} shows all three constraints on a single plot.
These constraints, in combination with the $\Delta\chi^2$ contours,
severely restrict the region of allowed solutions.

The uncorrected disk mass-to-light ratio of our formal best-fit solution is 
$(M/L_R)_d=3.9$, which is a bit on the high side.
The mean luminosity-weighted $B-R$ color of M31's disk
is $B-R\approx 1.60$ \citep{wal87}.  Using the color-$M/L$ relations of
\citet{bell01}, this corresponds to a mass-to-light ratio of $M/L_R \approx
3.5$.  Admittedly, the above $B-R$ color is uncorrected for reddening due to dust
while the \citet{bell01} relations are for dust-free quantities. However as they
note, the very dust that is responsible for the reddening also extinguishes the
light from the galaxy, increasing the $M/L$ ratio, and to first order the two
effects cancel out. A more careful analysis suggests that the $M/L$ derived
using uncorrected colors can be an overestimate by as much as $\sim$0.1 dex. We
therefore adopt a value of $M/L_R\approx 3.3$ (uncorrected) for the M31 disk,
though we recognize that this ratio could be as low as 2.8.  

We can also estimate the corrected or true mass-to-light ratio for the M31 disk.
The constant foreground reddening inside the Milky Way in the direction towards
M31 is $E(B-V)=0.08$ \citep{bur84} whereas the internal reddening in the disk of
M31 is estimated to be $E(B-V) \approx 0.25$ \citep{wil01} for a total reddening
of $E(B-V) \approx 0.33$.  Moreover, \citet{bar00} find that for the M31 disk (see
their Table 6), $E(B-R) \approx 1.6\times E(B-V) \approx 0.53$.  Correcting the
observed $B-R$ color of M31 disk by this amount and then, using the \citet{bell01}
correlations yields $M/L_R\approx$1.3.  Comparing this to the uncorrected
$M/L_R\approx 3.3$ implies that the extinction in $R$ is $\sim$ 1.0 mag, which is
in good agreement with the \citet{ken89} estimate of $\sim$0.99 mag for the
total extinction.  This agreement gives us confidence in our adopted value for
the uncorrected mass-to-light of the M31 disk.

The relationship between $(M/L_R)_d$ and the disk central surface density in our models
is $(M/L_R)_d=8.46 \Sigma_{O,9}$ where $\Sigma_{O,9}$ is $\Sigma_O$ in units of $10^9\; 
\msun / {\rm kpc}^2$.  In Figure~\ref{constrain}, we show the vertical line in the 
$\Sigma_0$--$r_h$ plane corresponding to $M/L_R=3.3$.  We assert that suitable 
M31 mass models lie on the segment of this line that falls within the existing 
$\Delta\chi^2 = 4.0$ confidence region. In Table~\ref{param}, we give the parameters for the 
best fit-mass model subject to the additional constraint that $(M/L_R)_d=3.3$.
We shall refer to this fit as the ``$M/L$-constrained'' solution,
and we shall use it in preference to the formal best-fit solution for the
remaining discussion.

Another interesting variant is denoted by the diamond located at the
intersection of the $C_{200} = 7$ and $f_{\rm gal} = 1$ constraints.  
We refer to this as the ``maximum disk'' solution  
and its properties are also listed in  Table~\ref{param}.  
This solution differs from the
traditional maximum disk models in that it
is required to be consistent with
the dynamical data {\it as well as} the physical constraints on the 
halo concentration parameter and baryonic fraction.  Due to the 
latter two restrictions, the disk in our ``maximum disk'' solution
does not dominate the gravitational potential to the same extent as
in the traditional ``maximum disk'' models. 
The disk mass-to-light ratio of our ``maximum disk'' model is 
$M/L_R=6.3$. 

Comparing these two constrained solutions to our unconstrained results, we find that the
bulge properties, such as the model mass as well as the realization of the bulge 
velocity dispersion and surface brightness profiles, are nearly
identical.  The altered surface density in the disk is offset by
the halo becoming denser and more concentrated for the constrained 
$(M/L_R)_d=3.3$ model, and the reverse for the maximum disk model,
as seen in Figure~\ref{rotplot} and Table~\ref{param}.   
The overall effect is that the total rotation profiles for the 
constrained and the unconstrained cases are nearly the same.  
In the $M/L$-constrained model, the halo contribution to the potential 
is greater than that of the disk at {\it all} radii, a property that according to 
\citet{wid03} and \cite{wid05} is likely to delay or even suppress the onset of 
bar formation. 
In contrast,   the gravitational contribution of the
disk in our maximum disk model is greater than that of the other two components
over the radial range 5 kpc $< r <$ 25 kpc, and based on their N-body simulation
studies, \citet{wid05} indicate that such  models are highly susceptible to
the formation of strong persistent bars.

\subsection{Comparing to Other Mass Models}\label{others}

\citet{ken89} was one of the first to construct a detailed mass model
for M31 using both photometric and kinematic data.  Comparing our $M/L$-constrained 
model (Table~\ref{param}) to 
his small bulge model, we find that his bulge properties are 
in excellent agreement with ours:
 Kent's  bulge is a factor of $\sim$ 1.2 
more massive than ours and 
the corresponding mass-to-light ratio is $(M/L_R)_b=(3.7\pm0.4)^1$, 
compared to our value of 3.9.
Kent's bulge is represented as an oblate spheroid whereas our bulge is 
spherically symmetric, while other differences in our assumptions are
detailed in \S~3.1 above.
In view of these differences in the models,
the level of agreement in our results for the M31 bulge
supports our prior 
assertions regarding the robustness of the bulge parameters.
Kent's disk, on the other hand,
is approximately twice as massive as that of our M/L-constrained model,
 and the corresponding mass-to-light ratio of (7.6$\pm$0.8)\footnote{These values are slightly 
different from those quoted in \citet{ken89} because they have been rescaled to account
for the differences in our assumed distances to M31, and 
converted from r-band to R-band using M31 colors and $(r-R)_\odot=0.18$
(http://www.sdss.org/dr4/algorithms/sdssUBVRITransform.html) }
is also larger than our value of 3.3.   
As indicated by his Figure 2, the disk is the primary contributor
to the  rotation profile over the range $5\,\kpc < r < 25\,\kpc$ whereas in our
preferred model, the disk does not dominate at any radii.   
However, the bulge and disk properties, including the mass-to-light ratios,
of Kent's model are similar to those of our maximum-disk model.
Among the differences in our two schemes are: (a) \citet{ken89} fixed
the central surface brightness of his disks to values lower than what we find; 
and (b) he adopted a constant density dark matter halo whereas we assume an 
NFW functional form.  The differences
in our preferred results seem primarily
due to assumption (b).  
Given these different assumptions underlying our two approaches,
and the disk-halo degeneracy discussed in \S~4.1,
the disagreement in our disk results, at a factor $\sim 2$ level,  
is not surprising.

\citet{wid03} developed their M31 model much like
we did; they sought out best-fit solutions by minimizing a composite
$\chi^2$ statistic, and some of the datasets we use are taken from
their paper.  However, their approach differs from ours in that they
modeled the disk, bulge and halo components of the galaxy as
distribution functions, which cannot be specified analytically.  
In practise, this meant that their halo was
represented by a lowered Evans model \citep{kuij94, eva93} instead of
an NFW-like profile, their bulge by a lowered isothermal sphere or a
King model \citep{king66, bin87}, and their disk by a
\citet{kuij94} model.  This resembles our disk in that its surface
density falls off exponentially with radius; 
unlike us, they also take into account the finite
thickness of the disk.  Widrow et al.\ list the properties of their best-fit
model as well as a number of variants.  In terms of its
characteristics for $r < 30$~kpc, our best-fit $M/L$-constrained 
model is similar to their preferred model
(Model A), as can be seen by comparing their Figure~4 to our
Figure~\ref{rotplot}.  There are, however, some quantitative
differences.  Their decomposition of the surface brightness into bulge
and disk components gives a bulge-to-disk light ratio of 0.58, versus
our ratio of 0.38.  This may stem from their different functional
forms for the disk and bulge, or it may be because they use only the
global surface brightness from \citet{wal87}, which does not constrain
the bulge region very well.  Their bulge-to-disk {\it mass} 
ratio, in contrast, 
is slightly smaller than ours: 0.37 versus 0.44.  As a result, 
their bulge $M/L$ is 2.7, 30\% smaller than ours, and their disk $M/L$ is 
4.4, 33\% larger than ours.
The contrast between their bulge and disk $M/L$ is not very
reasonable, given that the observed disk color is slightly bluer than
that of the bulge, unless there is substantial dark matter in the disk.
As for the halo, Widrow et al.~replace their lowered Evans model after
the fact with an NFW halo that closely matches the inner structure of
their original halo.  This NFW halo has $C_{200}=11.5$, $r_h=19.5$
kpc, $R_{200}=224$ kpc, and $M_{200}=1.3\times 10^{12}\; \msun$.  The
resulting total mass of M31 is roughly a factor of 1.8 greater than
that of our system, and is somewhat on the high side in light of the
mass constraints displayed in Figure~\ref{massplot}.  Nevertheless,
the disk and halo properties still put their model within the
allowed region in Figure~\ref{constrain}, and in general the
differences between their model and ours are not very substantial
given the large differences between our functional forms.

\citet{klyp02} also constructed a mass and light
model of M31, in order to test the consistency of large disk galaxies
such as the Milky Way and M31 with the cuspy halos of the LCDM
paradigm.  An interesting feature of this model is the inclusion of
adiabatic contraction \citep{blumenthal86} of their {\it initially
NFW} halos by the gravity of the disk and the bulge.  To fit the
baryonic component and the halo, they use a 4-component model
containing a flattened bulge, a bar, an exponential disk, and the
halo.  This contains about 14 parameters in total.  They do not
actually state the parameter values or how the parameters were chosen,
but they assert that these achieve a good match to the data without
the need for careful optimization.  Their treatment of the bulge and
disk dynamics is even more approximate than ours; the central mass
profile is calculated from an isothermal sphere approximation, and the
disk rotation curve apparently does not take flattening into
account.  For this reason, some differences from our results are to be
expected.  Nevertheless, we find a remarkable level of agreement
between their Model C1 and our best-fit constrained model, as can be
seen by comparing their Figure~4 to our Figure~\ref{rotplot}.
One primary difference is that their bulge-disk decomposition is
different, with the mass ratio between the two only 0.27 whereas in
our model it is 0.44.
This could be a product of the dynamics
modeling or of the different bulge profiles.  A second difference is
that, correcting their unextincted luminosity to extincted using the
transformations they suggest, they obtain a total galaxy R luminosity
of $3.5 \times 10^{10} {\rm L}_\odot$ or 16\% larger than ours;
we do not understand the origin of this
difference, since we are using nearly the same data and they have not
raised the \citet{wal87} surface brightness by 0.1 mag as we have.
Together with their total baryonic component mass,
which is smaller than ours by about 10\%, 
the higher luminosity implies a total galactic (bulge+disk)
mass-to-light ratio of $M/L_R = 2.5$
(extincted) whereas in our model it is 3.5.
A third difference is
that they use a somewhat more massive halo. Overall, we estimate that
the mass of their total system within 125 kpc is approximately
$0.9$--$1.0\times 10^{12}\;\msun$. 
In  our best-fit constrained
model, the total mass within 125 kpc is roughly a factor of 1.5 less.
This difference is
probably within the large errors on the halo mass and is not
necessarily a product of our different modeling technique.

\citet{ber01} and \citet{ber02} modeled the gravitational potential in
the inner $\sim 10$ kpc of M31, using a rotating triaxial bulge.  By
comparing to the molecular gas flows seen in high-resolution CO
velocity maps, they found a best-fit bulge mass of $2.3 \times 10^{10}
\msun$ within 3.5 kpc (see \citealt{ber02}).  In our three models in 
Table~\ref{param}, 
the bulge mass within this radius is $2.4 \times 10^{10} \msun$.
The agreement with the \citet{ber02} results is excellent,
especially considering the
complete independence of the datasets involved, and the simplifying
assumptions we have made in the treatment of the bulge which (as
emphasized above) is not the principal concern of our models.
\citet{ber01} and \citet{ber02} also estimate the disk and the
total mass within 3.5 kpc to be $1.2\times 10^{10} \msun$ and
$3.7\times 10^{10} \msun$, respectively.  Given that the latter 
is nearly the same as the combined disk+bulge contribution, 
\citet{ber02} argue that the gravitational contribution of the 
halo within 3.5 kpc must be negligible.  Our analysis, however,
does not support this latter conclusion.  We find that although
the  \citet{ber02} disk mass estimate is comparable to that in 
our formal best-fit and M/L-constrained models ($1.1$ and 
$1.0 \times 10^{10} \msun$, respectively), we require a total mass
within 3.5 kpc of $5 \times  10^{10} \msun$, a factor of
1.35 times larger, to account for the rotation curve data.  
Consequently, the dark halo contributes
roughly a third of the total gravity within 3.5 kpc.

In their analyses of the stream dynamics, \citet{iba04} considered 
two variants of the
\citet{klyp02} Model C1 in which the
bulge and disk properties were kept fixed to the values specified by Klypin et al.,
but the halo properties were allowed to vary with the aim of minimizing discrepancies
between the orbit of the progenitor and that of the giant stream.  
Depending on which stream constraints they used, 
\citet{iba04} iterated to two models.  The rotational velocity and the total 
mass profiles (computed from the rotation curve using the spherical approximation)
for these two models are shown as 
short dashed and dot-dashed (green)
lines in lower right panel of Figure~\ref{rotplot} and in Figure~\ref{massplot}.
As illustrated in Figure~\ref{rotplot}, the rotation curve
for the model dubbed ``high mass'' (short dashed)  by \citet{iba04} is in
onflict with the observations.  \citet{iba04} note this and reject the model.  
The characteristics of the alternate ``low mass'' model (dot-dashed curve)
are broadly similar
to those of our constrained best-fit axisymmetric case, and for the 
purposes of orbit calculations, we expect that the two will give broadly 
similar results.  However, the fact that the \citet{iba04} system has 20\% 
more mass within 125 kpc will, for a given set of initial conditions,
lead to small differences in the apocenter.  
We discuss this point further in the next section.

\citet{fon04}, in their recent study,  adopt the axisymmetric bulge-disk-halo
M31 mass model of \citet{bek01} with potential:
\begin{equation}
\Phi(r)=\Phi_b(r)+\Phi_d(r)+\Phi_h(r),
\label{bekki}
\end{equation}
where,
\begin{align}
\Phi_{b}(r) &=-\frac{G M_{b}}{r_{b}+r}, \\
\Phi_d(r)   &=-\frac{G M_d}{\sqrt{R^2+(a+\sqrt{z^2 + b^2})^2}},\\
\Phi_h(r)   &= \frac{1}{2}{V_{h}}^2 \ln({r_c}^2+r^2),
\end{align}
$r$ is the distance from the center of the galaxy, $R$ and $z$ are polar
coordinates aligned with M31's disk plane, $r_c=12$ kpc, $V_h=186$
km s$^{-1}$, $M_d=1.3\times 10^{11}\;\msun$, $a=6.5$ kpc, $b=0.26$ kpc, 
$M_b=9.2\times 10^{10}\;\msun$, and $r_b=0.7$ kpc.   For comparative purposes,
the rotation curve and the total mass profiles (computed from the rotation curve
using the spherical approximation) for this model are plotted in the lower
right panel of Figure~\ref{rotplot} and in Figure~\ref{massplot} as 
long-dashed (cyan) curves.
The model rotation curve is systematically higher than the observed profile,
with the two being least discrepant in the outer disk region where the 
model curve is approximately $20$ km~s$^{-1}$ higher, and most discrepant
in the inner disk.  The latter is primarily due to the large bulge mass 
assumed by \citet{bek01}.  Their bulge mass is a factor $\sim 2.5$ larger 
than ours.  Additionally, the Bekki et al.~model overestimates the 
the system mass beyond $\sim$ 80 kpc.  On the whole, the Bekki et al.~model
describes a galaxy with a much deeper potential than that 
indicated by the observations or that described by our preferred M31 model.
As we shall illustrate in the next section, this has significant implications 
for orbit calculations.

To summarize, the models of the M31 potential presented in this paper seem to be more 
consistent with the observations than several others presented in the literature, and 
generally consistent with others of greater complexity while being easier to use for 
calculating orbits.  There is, however, still a great deal of uncertainty in the 
actual potential.
Our model, for example, is more strongly constrained in some ways
by consideration of the disk mass-to-light ratio, 
the halo concentration, and the galaxy baryonic fraction, than by the actual observations.
This uncertainty has bearing on studies of the response of the disk to perturbations,
the response of dark halos to the assembly of galaxies, and the orbits of satellite 
galaxies, among other issues. We expect that the model can be improved in the future 
by the addition of more halo tracers, such as the halo stars detected by \citet{guh05}
and Chapman et al.~(in preparation); by using coherent stellar streams, such as the giant
southern stream and the stream in the vicinity of NGC 205 \citep{mcc04};
by the use of deep infrared surface brightness measurements, which 
are less sensitive to dust and stellar population variations than 
the optical data used here; by the incorporation of non-axisymmetric
features such as the oblong/triaxial bulge and the spiral features 
and warping of the disk; and by properly treating the finite 
thickness of the disk.

\section{Sample Orbits in Spherical and Flattened Potentials}\label{obscon}

With our mass models defined, we are now in the position to study the
orbits giving rise to the giant southern stream in M31.  The results
of our study are described in detail in Paper II \citep{paperII}.
Here, we merely present a few examples that illustrate 
the sensitivity of the orbits to the mass models used. 

To facilitate comparison, we start by adopting initial conditions
similar to those used by \citet{fon04}.  
These authors compute an orbit that is
in rough agreement with the kinematic observations of the giant
southern stream by \citet{mcc03}, \citet{law05}, \citet{iba04}, and
\citet{guh04}.  We assume here, with these authors, that the stream and the
progenitor orbit are coincident.  (This condition will be 
modified in Paper II).  Font et al.\ define a coordinate
system ($x,y,z$) such that the $x$--$y$ plane is in the sky,
with $x$ increasing to the E and $y$ to the N, and $z$ is
parallel to the line of sight.  
Font et al.\ locate the starting
point of their orbit in the centre of the southern stream.  For a
detailed discussion of the stream geometry and of the choice of
initial conditions, we refer the reader to \citet{fon04} and our
Paper II.  

We slightly optimize the Font et al.~initial 
conditions to better match the observations of the stream in our
$M/L$-constrained disk potential.  Specifically, we set 

\noindent
\begin{tabular}{ l  l }
\( x_0=   \phantom{-}14.2~{\rm kpc},\)        &  \(v_{x0}=-\phantom{1}50\phantom{.02}~{\rm km~s}^{-1},\)      \\  
\(y_0=   -33.2~{\rm kpc},\)                   &  \(v_{y0}=          \phantom{-1}86.02~{\rm km~s}^{-1},\)    \\    
\(z_0=  \phantom{-}60\phantom{.2}~{\rm kpc},\) &  \( v_{z0}=          -158\phantom{.02}~{\rm km~s}^{-1}.\) \\  
\end{tabular}

For the comparisons here, the details 
of this optimization procedure are not important.
Issues surrounding the
comparison to observations will be discussed in Paper II.
The primary thing to note is that the apocenter of
the orbit must occur approximately at Field 1, since the $z$ velocity
of the stream relative to M31 is nearly zero there.  

\begin{figure}
\begin{center}
\includegraphics[width=84mm,bb=0 5 252 576]{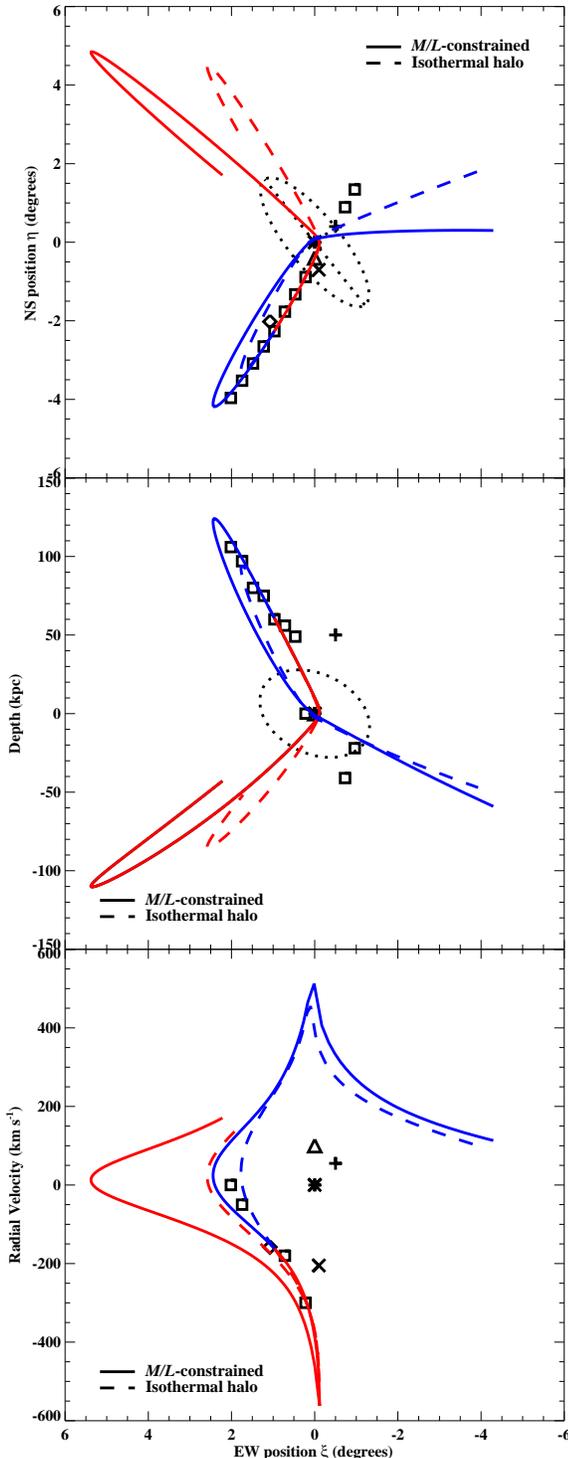}
\caption{
Comparison of test particle orbits with exactly the
same initial conditions but in two different 
potentials: the simple singular isothermal sphere potential 
discussed in the text (dashed 
curve) and the potential corresponding to our $M/L$-constrained model
(solid curve).  The orbits are started in the centre of the southern 
stream with initial velocity  such that the orbit in our
potential turns around near Field 1.  
Red and blue 
curves represent orbit integrations forward and 
backward in time, respectively.
(Top panel): shows the orbit in sky
coordinates, $\xi$ and $\eta$.
(Middle panel): plots the line-of-sight depth (in kpc) 
of the orbit versus the $\xi$ position in the sky. 
(Bottom panel): plots the variation of radial velocity (in km $s^{-1}$) 
with $\xi$ position in the sky. 
The symbols in the three
panels are as follows: star (M31), triangle (M32), plus (NGC
205), cross (And VIII), diamond (Field a3), squares 
(\citealp{mcc03} stream fields). 
}
\label{orbit1}
\end{center}
\end{figure}

The main difference between the initial conditions that we 
use here and those used by Font et~al.\ is that our
initial velocity is not as high as theirs, because our
potential for M31 is not as deep as the one they use.
As a simple illustration of this point, Figure~\ref{orbit1} 
shows test particle orbits
calculated forward and backward from our
new initial conditions, but computed in two
different potentials: our $M/L$-constrained mass model
(solid curve) and the simple singular isothermal sphere halo
(dashed curve) 
with potential given by 
\begin{equation}
\Phi(r)={V_{c}}^2 \ln{(r)},
\label{fontpot}
\end{equation}
where $V_{c}=200$~km~s$^{-1}$ is the circular velocity.  
The latter is a good approximation to the model used by
Font et al.\ in the outer halo.  The key thing to note in
Figure~\ref{orbit1} is that in the deeper singular isothermal 
potential, the test particle is unable to climb out as far before 
reaching apocenter when its orbit is traced backward in time, and
it falls short of Field 1 by approximately one degree.  To guarantee
that the apocenter of the test particle's orbit falls in Field
1 requires endowing the test particle with higher initial velocities.   
Since the radial velocities along the stream are observed quantities,
the differences in the initial velocities will eventually result 
in a different assumed tilt of the orbital plane, and thus different 
projections of subsequent lobes of the orbit.

\begin{figure}
\begin{center}
\includegraphics[width=84mm]{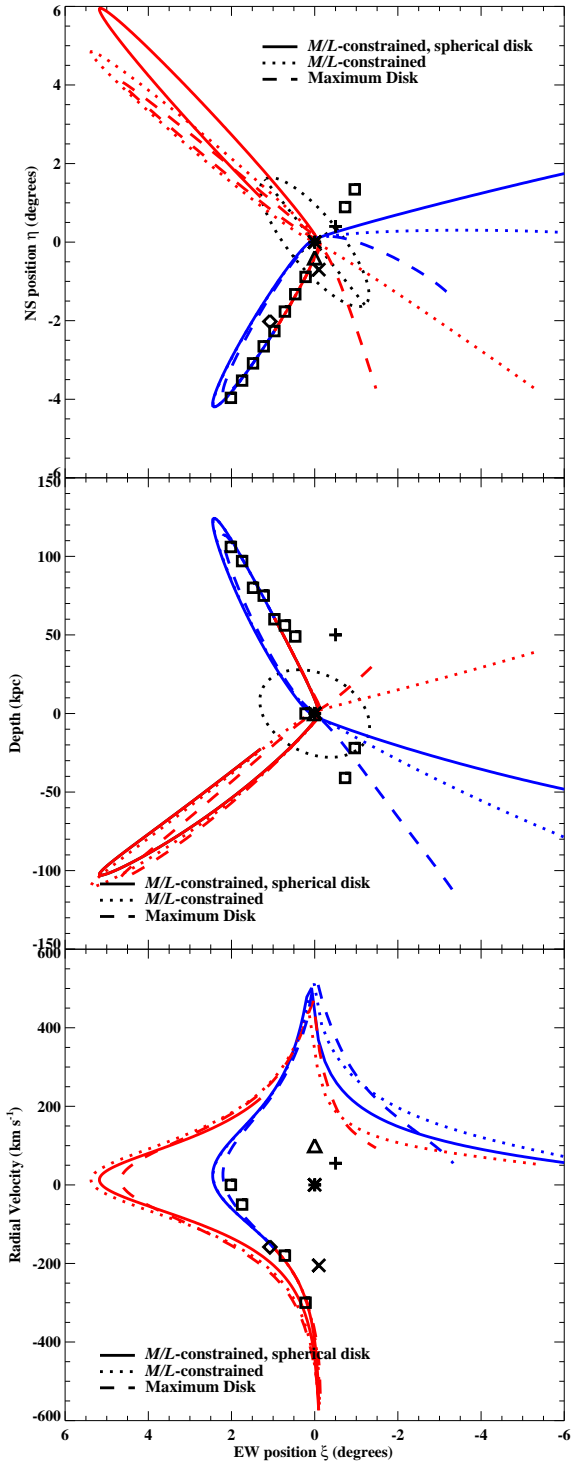}
\caption{The above three panels show a comparison between the models
with different strengths of the disk potential.  All panels, symbols,
and colours are the same as in Figure~\ref{orbit1}; however, the 
dashed line now represents the ``maximum disk'' model, the dotted line
represents the $M/L$-constrained model, and the solid line represents
the $M/L$-constrained 
``spherical disk''
model (see text for details).
All orbits are calculated using the
same initial conditions as in Figure~\ref{orbit1}.  
Note that the time intervals
for the different orbits are not the same.}
\label{orbit2}
\end{center}
\end{figure}

In Figure~\ref{orbit2}, we examine the effect of the strength of the
flattened disk potential.  The dotted line shows the same results for
the $M/L$-constrained model as in Figure~\ref{orbit1}.
The solid line shows the results for the {\em spherical disk mass model} with 
$M/L$-constrained model parameters. 
(We remind the reader that the halo and the bulge components of these
two models are exactly the same, and even the radial mass profiles of
the disk component $M_d(<r)$ for the two models are the same.   The only
difference is that in the spherical disk case, the disk mass is distributed 
in a spherically symmetric fashion, as opposed to planar and axisymmetric.
Consequently, the $M/L$-constrained spherical disk model is, as the name suggests, a 
spherically symmetric analog of our axisymmetric $M/L$-constrained model.)
The result for the spherical disk model is essentially
indistinguishable from the flattened disk model over the duration of the
radial lobe that includes the stream.  However, the directions of the
preceding and subsequent lobes are different.  
These parts of the trajectories are very sensitive to the differences
in the geometry of the disk potential at pericenter, since the test
particle plunges all the way in to where the disk contribution to the
total potential is significant.  The gravitational pull from the disk,
pulling the test particle towards the plane of the disk, results in an
orbit that bends more strongly than in the spherical case.
The dashed line uses the results from our ``maximum disk'' model.
Again, the primary effect is a change in the direction of the 
previous and subsequent radial loops 
relative to the orbit in our fiducial $M/L$-constrained potential
although in this case, the slightly deeper potential also results
in a slightly smaller radial excursion at apocenter.
Both of these effects are due to the stronger gravity of the disk in the 
maximum-disk model.

In summary, the radial mass profile has a significant effect on orbits
in the halo, and is the primary factor in determining the length of
the stream given the initial velocity, or alternatively the velocity
gradient given the length of the stream. The geometric shape of the
mass profile is a less significant factor, at least with the tests
performed here, but it does significantly affect the direction of
subsequent lobes.  If part of the objective for computing such orbits
is to try to ascertain where the progenitor might be, then ensuring
that the orbits are being computed in a realistically shaped potential
is important.  We have neglected the possible flattening of the halo,
both here and in Paper II, but this is clearly an interesting
direction to explore at some point.

\section{Summary}\label{summary}

The following are several significant results from this work that are
important to highlight:

\begin{itemize}
\item We have presented a simple analytic model for the mass density,
light density, and gravitational potential of M31, which is
easy to use for the purposes of orbit calculations.  We apply this
potential in Paper II to the problem of estimating
the orbit and other properties of the giant southern stream in M31's
halo.

\item Our new potential does a better job matching the dynamics of
M31 than other simple potentials used recently
for the dynamics of the stream.  It does a comparable job,
and in fact is fairly similar in its properties, to several more 
sophisticated but non-analytic models in the literature.

\item The eight parameters in our mass and light model span a
large but low-dimensional region in parameter space.
Physical constraints on the fraction of halo baryons in the galaxy,
the halo concentration, and the stellar mass-to-light ratio
restrict the parameters to a much smaller region.  
Our preferred model uses the $M/L$ ratio of the disk 
as a further constraint.
The parameters of this model and its important physical 
quantities are summarized in Table~\ref{param}.

\item The radial mass profile has a significant effect on 
orbits within the giant southern stream, and is 
the determining factor in the length of the stream for a given initial
velocity.  The geometric shape of the mass profile does not play as
significant a role in determining the orbit as the radial mass profile,
at least within the scope of possible variations of the disk surface density.
However, it does have an effect on the direction of subsequent lobes
of the orbit.  This becomes important if one wishes to construct
orbits for the purposes of locating the progenitor of the stream.
\end{itemize}

\vskip+2mm
We are deeply grateful to Larry Widrow for kindly providing us with his data and 
the results of his mass model for M31.   We further thank him as well as 
Rene Walterbos, Mark Wilkinson, Chris Pritchet, Mike Rich, Sandy Faber,
Martin Weinberg, Andreea Font, Kathryn Johnston, and Laurent Loinard
for helpful advice and clarifications.   
We are also indebted to the
anonymous referee for most useful recommendations.
Research support for JJG, MF and AB comes from the Natural Sciences and 
Engineering Research Council (Canada) through the Discovery and the 
Collaborative Research Opportunities grants.  AB would also like to 
acknowledge support from the Leverhulme Trust (UK) in the form of the
Leverhulme Visiting Professorship.  PG is supported by 
NSF grant AST-0307966. He is grateful to the HIA/DAO/NRC staff for 
graciously hosting his 2002-03 Herzberg fellowship during which time 
this collaborative project was conceived.

\label{lastpage}

\end{document}